\documentclass{ieeetmlcn}
\usepackage{cite}
\usepackage{amsmath,amssymb,amsfonts}
\usepackage{graphicx,color}
\usepackage{textcomp}
\usepackage{xcolor}
\usepackage{hyperref}
\hypersetup{hidelinks}
\usepackage{algorithm,algorithmic}
\usepackage[caption=false]{subfig}
\usepackage{threeparttable}
\usepackage{url}
\usepackage{cite}
\DeclareMathAlphabet{\pazocal}{OMS}{zplm}{m}{n}
\usepackage{nicefrac}
\usepackage{optidef}
\usepackage{gensymb}
\usepackage{hyperref}
\usepackage{xurl}
\usepackage{orcidlink}
\usepackage{booktabs}
\usepackage{xcolor}
\usepackage[english]{babel}
\usepackage{soul,color}
\soulregister\cite7
\soulregister\ref7
\soulregister\ac7
\soulregister\acp7
\soulregister\underline7

\def\BibTeX{{\rm B\kern-.05em{\sc i\kern-.025em b}\kern-.08em
    T\kern-.1667em\lower.7ex\hbox{E}\kern-.125emX}}
\AtBeginDocument{\definecolor{tmlcncolor}{cmyk}{0.93,0.59,0.15,0.02}\definecolor{NavyBlue}{RGB}{0,86,125}}

\def\OJlogo{\vspace{-4pt}\includegraphics[height=18pt]{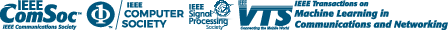}}
\def\seclogo{\vspace{10pt}\includegraphics[height=18pt]{new_logo}}

\def\authorrefmark#1{\ensuremath{^{\textbf{#1}}}}

\begin{document}
\receiveddate{XX Month, XXXX}
\reviseddate{XX Month, XXXX}
\accepteddate{XX Month, XXXX}
\publisheddate{XX Month, XXXX}
\currentdate{XX Month, XXXX}
\doiinfo{TMLCN.2022.1234567}

\markboth{Signal Whisperers: Enhancing Wireless Reception Using DRL-Guided Reflector Arrays}{Hieu Le {et al.}}

\title{Signal Whisperers: Enhancing Wireless Reception Using DRL-Guided Reflector Arrays}

\author{Hieu Le \authorrefmark{1} \orcidlink{0000-0003-2510-073X}, Oguz Bedir \authorrefmark{1} \orcidlink{0000-0003-2871-0437}, Mostafa Ibrahim \authorrefmark{2} \orcidlink{0000-0003-0507-6752}, Jian Tao \authorrefmark{1,3} \orcidlink{0000-0003-1374-8678}, and Sabit Ekin \authorrefmark{1,2} \orcidlink{0000-0002-9957-7752}}
\affil{Electrical and Computer Engineering, Texas A\&M University, College Station, TX 77840 USA}
\affil{Engineering Technology and Industrial Distribution, Texas A\&M University, College Station, TX 77840 USA}
\affil{School of Performance, Visualization, and Fine Arts, Texas A\&M University, College Station, TX 77840 USA}
\corresp{Corresponding author: Hieu Le (email: hieult@tamu.edu).}

\begin{abstract}
{This paper presents a multi-agent reinforcement learning (MARL) approach for controlling adjustable metallic reflector arrays to enhance wireless signal reception in non-line-of-sight (NLOS) scenarios. Unlike conventional reconfigurable intelligent surfaces (RIS) that require complex channel estimation, our system employs a centralized training with decentralized execution (CTDE) paradigm where individual agents corresponding to reflector segments autonomously optimize reflector element orientation in three-dimensional space using spatial intelligence based on user location information. Through extensive ray-tracing simulations with dynamic user mobility, the proposed multi-agent beam-focusing framework demonstrates substantial performance improvements over single-agent reinforcement learning baselines, while maintaining rapid adaptation to user movement within one simulation step. Comprehensive evaluation across varying user densities and reflector configurations validates system scalability and robustness. The results demonstrate the potential of learning-based approaches for adaptive wireless propagation control.}
\end{abstract}



\maketitle

\section{INTRODUCTION}

\IEEEPARstart{T}{he} {exponential growth in wireless data traffic, driven by emerging applications such as augmented reality, autonomous vehicles, and Internet of Things deployments, has pushed traditional wireless communication systems to their fundamental limits. Conventional approaches treat the wireless propagation environment as a static, uncontrollable obstacle that impedes signal transmission, requiring increasingly sophisticated signal processing techniques and higher transmission powers to overcome channel impairments. This paradigm has led to diminishing returns in system performance while simultaneously increasing energy consumption and infrastructure complexity \cite{direnzo:2020}.

Reconfigurable intelligent surfaces (RIS) represent a paradigm shift toward programmable wireless environments, where passive objects become active participants in signal propagation control. However, despite their theoretical promise, practical RIS implementations face several critical deployment barriers. The most significant challenge is the overwhelming complexity of channel state information (CSI) estimation, requiring precise characterization of electromagnetic properties at hundreds or thousands of reflecting elements simultaneously. This creates computational overhead that scales exponentially with system complexity \cite{bjornson2022reconfigurable}.

Furthermore, existing RIS approaches rely heavily on constructive interference optimization, demanding sophisticated hardware capabilities including high-resolution phase shifters and low-latency reconfiguration mechanisms. These stringent requirements, combined with the need for perfect synchronization across all reflecting elements, have prevented widespread commercial deployment \cite{pan2022overview, kim2022practical, a9864655}

This work introduces a fundamentally different approach that eliminates CSI estimation entirely by operating at a higher abstraction level. Instead of orchestrating precise electromagnetic interference patterns, our method leverages user location information to optimize reflection geometries through large-scale propagation control.

The key innovation lies in formulating the reflector control problem as a cooperative multi-agent reinforcement learning (MARL) problem, where individual agents corresponding to reflector segments learn to cooperate autonomously through a centralized training with decentralized execution (CTDE) paradigm. This approach decomposes the complex system-wide optimization into manageable sub-problems, enabling scalable and adaptive operation without requiring explicit coordination protocols or information exchange between reflecting elements. Moreover, MARL is well-suited for addressing non-convex, high-dimensional, and partially observable optimization problems, while maintaining adaptability to dynamic environments through continual learning. Additionally, practical deployment feasibility is enhanced by the widespread availability of neural network processing hardware and specialized deep learning accelerators in modern communication systems \cite{nasari2022benchmarking, le2024insight, qualcomm2024unlocking}.

Unlike conventional RIS systems that rely on passive metamaterial elements, our approach employs mechanically adjustable metallic reflectors that dynamically modify their orientation. While introducing mechanical complexity, our design choice offers several advantages: (1) broader frequency operation ranges (by exploiting frequency-independent geometric reflection principles rather than relying on frequency-specific electronic circuits and metamaterial resonances), and (2) simplified control mechanisms using commercial off-the-shelf servo systems. 

This work makes three primary contributions:

}
\begin{itemize}
    \item {\textbf{CSI-free operation}: We formulate reflector control as a multi-agent Markov Decision Process (MA-MDP), enabling deep reinforcement learning techniques for wireless propagation control without requiring CSI estimation.}
    
    \item {\textbf{Substantial performance gains}: Through extensive ray-tracing simulations in an L-shaped hallway environment, we demonstrate up to 20.7 dB improvement over static flat reflectors.}
    
    \item {\textbf{Comprehensive validation}: We establish practical viability through performance validation across varying user densities, reflector configurations, and different degrees of error in user position information, demonstrating system robustness and scalability.}

\end{itemize}

{
By eliminating the fundamental barriers preventing practical RIS deployment, this work paves the way for autonomous smart radio environments that can adapt to real-world complexities while maintaining economic feasibility. Comprehensive robustness analysis under realistic positioning uncertainties confirms that the system maintains performance advantages even with localization accuracy limitations, addressing a critical practical deployment concern. The approach is particularly relevant for applications requiring reliable coverage in challenging propagation environments such as indoor millimeter waves (mmWave) systems. Our work is publicly available at} \url{https://github.com/hieutrungle/saris_revised}.

The rest of our study is structured as follows: Section~\ref{sec:review} provides a literature review. Section~\ref{sec:system_model} presents the system model and problem formulation. Section~\ref{sec:marl_ctde} describes the MARL framework. Section~\ref{sec:results_and_discussion} analyzes simulation results. Finally, Section~\ref{sec:conclusion} concludes with key findings and future directions.

\section{LITERATURE REVIEW}
\label{sec:review}

{
The landscape of intelligent technology has evolved rapidly over the past decade, driven by the exponential growth in wireless data demands and the fundamental limitations of conventional communication paradigms. Traditional RIS have emerged as a transformative technology, primarily focusing on orchestrating constructive interference at receivers through precise electronic phase manipulation to maximize achievable data rates \cite{direnzo:2020, zahra:2021}. However, these approaches fundamentally depend on accurate CSI at each reflecting element, a requirement that scales exponentially in complexity with system size. For implementations incorporating hundreds or thousands of elements with potential dual polarization capabilities, the CSI estimation burden becomes the primary bottleneck limiting practical deployment \cite{basharat:2022}.

Recent comprehensive studies demonstrate that obtaining reliable CSI in RIS-enabled systems requires complex procedures involving multiple sequential RIS configurations, with pilot overhead scaling as the product of reflecting elements and served users. Typical cascaded channel estimation methods result in pilot overhead reaching hundreds to thousands of symbols, creating prohibitive spectral efficiency losses and channel estimation delays that exceed coherence times in mobile environments \cite{a9400843}. Multiple sophisticated approaches have attempted to address this CSI bottleneck, including ON/OFF-based protocols with successive element activation \cite{a10053657}, discrete Fourier transform methods with optimized phase patterns \cite{a9328501}, and compressive sensing techniques exploiting mmWave channel sparsity \cite{a10016718}. Despite these advances, fundamental challenges persist: high-dimensional cascaded channel structures create ill-conditioned estimation problems, substantial pilot overhead scales unfavorably with system size, and hardware limitations compound estimation difficulties

Statistical CSI-based RIS have demonstrated that effective beamforming can be achieved without per-element instantaneous channel estimation. Multiport network theory approaches treat the RIS as a reciprocal multi-port scattering network, leveraging statistical channel state information to optimize average rate through eigenmode selection and covariance-aware phase alignment \cite{a10666709}. Blind beamforming methods eliminate CSI by iteratively adjusting phase shifts based solely on received power measurements, exploiting the law of large numbers to converge to constructive configurations and guaranteeing $O(N^2)$ signal-to-noise ratio boost for N-element arrays under rich-scattering conditions \cite{lai2023blind}. Codebook-based strategies extend these concepts by precomputing finite discrete phase-shift sets, with appropriately designed codebooks achieving near-full-CSI performance while reducing feedback requirements by orders of magnitude \cite{a9952197}. Location-based RIS techniques complement statistical methods by using coarse user position information rather than full CSI for phase design, with joint localization and beamforming frameworks optimizing Cramér-Rao lower bound-derived objectives to maximize received power at known coordinates \cite{nazar2024revolutionizing}.

The emergence of deep reinforcement learning (DRL) for RIS optimization has attempted to circumvent traditional limitations while introducing new complexities. Pioneering work by \cite{huang2020reconfigurable} demonstrated sum-rate maximization using full CSI knowledge across all communication links, while subsequent studies \cite{taha2020deep, taha2021enabling} developed DRL techniques for phase shift optimization using sampled channel vectors. In \cite{choi2024deep}, authors integrated sensing elements into RIS for distributed channel estimation and subsequent DRL optimization, achieving improved sum-rate performance but still requiring partial channel knowledge at the reflecting surface. However, these approaches still require channel estimation at the RIS, substantially increasing system complexity and energy consumption. Recent attempts to eliminate CSI dependence have shown mixed results, with deep learning schemes relating phase shifts directly to receiver locations requiring extensive offline training data that limits adaptability to dynamic scenarios \cite{sheen2021deep}.

The evolution toward MARL for RIS optimization represents significant advancement in addressing coordination challenges inherent in large-scale reflecting element control. 
Recent MARL implementations have demonstrated substantial improvements, with Multi-Agent Twin Delayed Deep Deterministic Policy Gradient (TD3) applied to joint beamforming and RIS codebook design achieving performance comparable to 256-beam Discrete Fourier Transform codebooks while reducing training overhead by 97\% \cite{a10060056}. The MAGAR framework for STAR-RIS systems demonstrates 18\% energy efficiency improvements compared to baseline MARL approaches \cite{a10758034}, validating distributed learning paradigms for intelligent surface control.

MARL algorithms can be categorized into three main groups based on the types of settings they address: fully cooperative, fully competitive, and mixed settings. In the cooperative setting, agents collaborate to optimize a common long-term return, where all agents usually share a common reward function~\cite{busoniu2008comprehensive, zhang2018fully}. This includes MA-MDP and more general models like team-average reward scenarios where agents may have different reward functions but work toward optimizing the average reward across all agents~\cite{kar2013cal, zhang2018fully}. The competitive setting is typically modeled as zero-sum Markov games, where the rewards of agents sum to zero, with most literature focusing on two-player scenarios for computational tractability~\cite{littman1994markov, shapley1953stochastic, park2023multi}. These settings also serve as models for robust learning by treating uncertainty as a fictitious opponent. The mixed setting, also known as general-sum games, imposes no restrictions on agent relationships and includes scenarios with both cooperative and competitive elements, such as teams competing against each other~\cite{hu2003nash, lowe2017multi}.

The CTDE paradigm has emerged as the dominant architectural approach in MARL-based RIS optimization, enabling systems to leverage global information during training while maintaining practical deployment constraints through local observation-based execution. 
Recent studies have validated CTDE effectiveness across diverse scenarios, with Multi-Agent Deep Deterministic Policy Gradient (DDPG) demonstrating 11\% improvements in task completion rates in vehicular networks \cite{a10261304} and modified MADDPG algorithms achieving 25\% cost reductions in vehicular edge computing \cite{a10654286}. Algorithmic innovations continue to emerge, with fuzzy-logic-enhanced MADDPG achieving 30\% faster convergence \cite{a10508095} and stochastic learning approaches reducing outage probability by 40\% under random blockage conditions \cite{a9314027}.

Despite extensive research in electronic RIS systems, mechanically adjustable metallic reflectors represent a fundamentally different paradigm that has received limited attention. Unlike electronic RIS systems requiring complex RF circuits, phase shifters, and sophisticated control mechanisms, metallic reflectors offer inherent advantages including broadband operation, simplified control through standard servo systems, and elimination of complex electromagnetic precision requirements. Recent studies have investigated passive metallic reflectors across various geometries at 28 GHz, revealing that strategically positioned flat reflectors achieve significant gain improvements in non-line-of-sight scenarios without electronic complexity \cite{a8972365, a9500547, le2024guiding}. In \cite{a10279522}, authors advanced theoretical understanding by developing mathematical models that refine conventional reflection principles, demonstrating limitations of traditional Snell's Law assumptions when reflector dimensions approach signal wavelengths.

The application of cooperative MARL to mechanically adjustable metallic reflector control represents an unexplored frontier addressing gaps in current intelligent surface technology. 
This paradigmatic departure offers the potential to eliminate the fundamental CSI bottleneck that has limited practical RIS deployment while maintaining significant performance improvements through large-scale propagation control rather than electromagnetic precision. Our work bridges this gap by introducing a fundamentally different approach that eliminates CSI estimation through spatial intelligence rather than electromagnetic precision, leveraging user location information to optimize reflection geometries at a higher abstraction level while enabling multi-agent coordination benefits demonstrated in RIS research.
}

\section{SYSTEM MODEL AND PROBLEM FORMULATION}
\label{sec:system_model}

\begin{figure}[!tbp]
  \centering
  \includegraphics[width=0.8\linewidth]{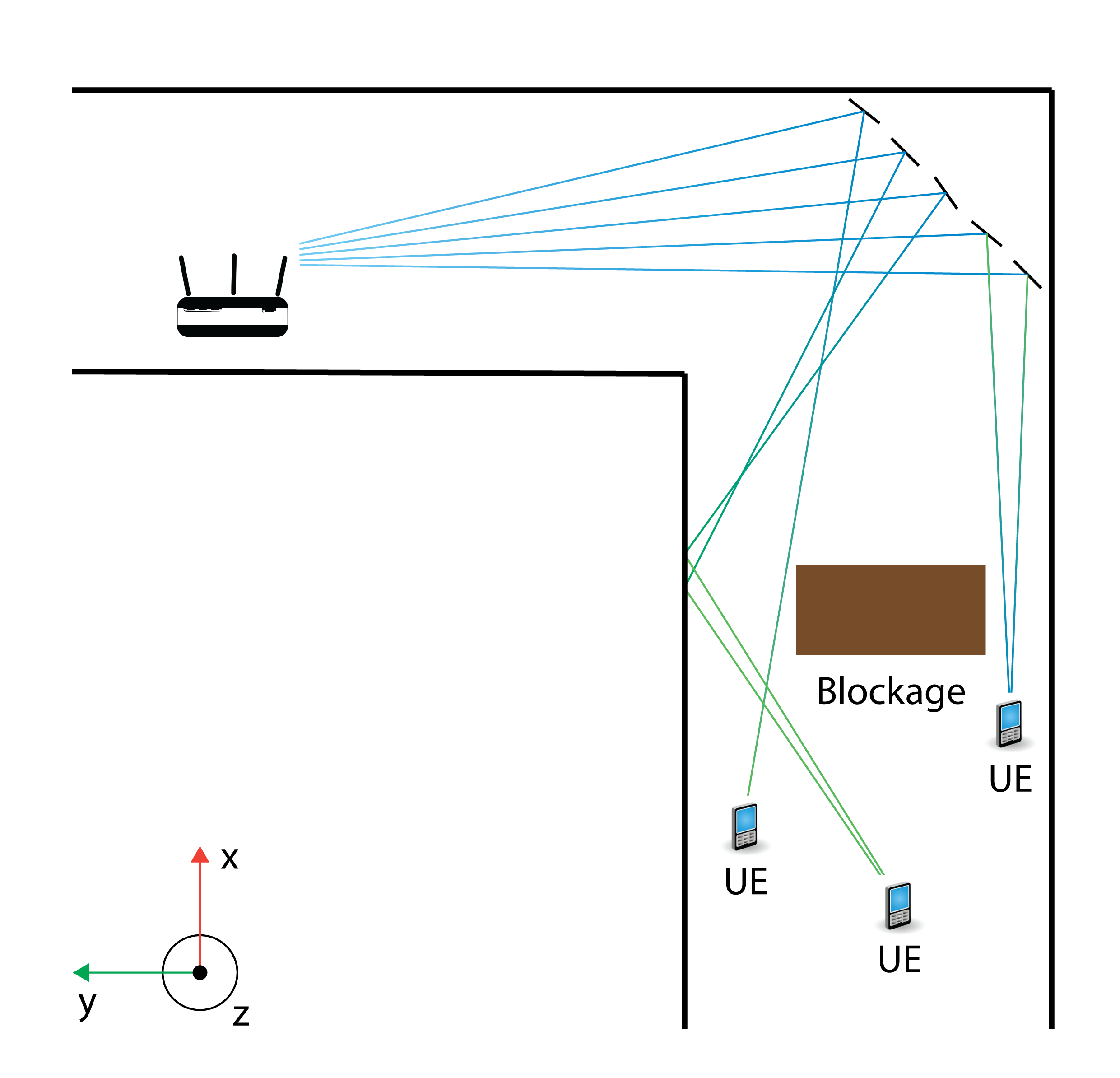}
  \caption{An example of an optimized reflector configuration to enhance signal quality for all UEs.}
  \label{figure:idea}
\end{figure}

\subsection{Formal Multi-Agent MDP Definition}

We consider a multi-user multiple-input multiple-output (MU-MIMO) wireless communication system enhanced by an intelligent reflecting surface comprising $N$ mechanically adjustable metallic reflector elements serving $K$ users in a complex propagation environment. Our system employs physically reconfigurable reflector tiles that can dynamically adjust their spatial orientation to optimize signal propagation characteristics. The system operates in environments characterized by obstacles that create heterogeneous channel conditions, where users experience varying combinations of line-of-sight (LOS) and non-line-of-sight (NLOS) propagation paths (Fig. \ref{figure:idea}). This complexity requires an adaptive control framework capable of autonomous operation while maintaining compatibility with existing wireless communication protocols.  

\subsubsection{Multi-Agent Markov Decision Process Formulation}

We formulate the reflector control problem as a MA-MDP defined by the tuple:
\begin{equation}
\mathcal{M} = (\mathcal{L}, \mathcal{S}, \mathcal{A}, \mathcal{P}, \mathcal{R}, \gamma),
\end{equation}
where:
\begin{itemize}
    \item $\mathcal{L} = \{1, 2, \dots, L\}$ represents the set of intelligent agents, with each agent $l \in \mathcal{L}$ corresponding to a group of reflector elements serving a specific user.
    \item $\mathcal{S}$ denotes the global state space encompassing all feasible system configurations.
    \item $\mathcal{A} = \mathcal{A}_1 \times \mathcal{A}_2 \times \dots \times \mathcal{A}_L$ is the joint action space as the Cartesian product of individual agent action spaces.
    \item $\mathcal{P}: \mathcal{S} \times \mathcal{A} \times \mathcal{S} \rightarrow [0,1]$ defines the state transition probability function.
    \item $\mathcal{R}: \mathcal{S} \times \mathcal{A} \rightarrow \mathbb{R}$ specifies the reward function.
    \item $\gamma \in (0,1]$ is the discount factor balancing immediate and future rewards.
\end{itemize}

\subsubsection{{State Space Definition}}

The global state at time step $t$ is defined as:
\begin{align}
    s_t = \big[ & u_{1,t}, \dots, u_{K,t}, r_{1,t}, \dots, r_{L,t}, \nonumber \\
                & f_{1,t}, \dots, f_{L,t} \big] \in \mathcal{S},
\end{align}
where $u_{k,t} = [u_{k,x,t}, u_{k,y,t}, u_{k,z,t}]^\top \in \mathbb{R}^3$
is the three-dimensional position of user $k$, and 
$f_{l,t} = [f_{l,x,t}, f_{l,y,t}, f_{l,z,t}]^\top \in \mathbb{R}^3$
denotes the controllable focal point for agent $l$. $r_{l,t} \in \mathbb{R}^3$ denotes the fixed position of the reflector segment controlled by agent $l$.

Each agent $l$ maintains a local observation:
\begin{equation}
o_{l,t} = [u_{\pi(l),t}, r_{l,t}, f_{l,t}]^\top \in \mathcal{O}_i,
\end{equation}
where $u_{\pi(l),t}$ is the position of the user assigned to agent $l$ via mapping $\pi: \mathcal{L} \rightarrow \{1,2,\dots,K\}$.

\subsubsection{{Action Space Formulation}}

The action space for each agent $l$ is:
\begin{equation}
\mathcal{A}_l = \{a_l \in \mathbb{R}^3 : \|a_l\|_\infty \leq \delta_{\max}\},
\end{equation}
where $a_{l,t} = [\Delta f_{l,x,t}, \Delta f_{l,y,t}, \Delta f_{l,z,t}]^\top$ represents the focal point displacement vector, and $\delta_{\max}$ defines the maximum allowable displacement.  

The joint action vector is:
\[
a_t = [a_{1,t}, a_{2,t}, \dots, a_{L,t}]^\top \in \mathcal{A}.
\]

\subsubsection{Physical Constraint Mapping}

For reflector element $(i,j)$ belonging to agent $l$, elevation $\theta_{i,j,t}$ and azimuth $\phi_{i,j,t}$ can be calculated using a bisector vector as:

\begin{equation}
    \overrightarrow{n_{i,j,t}} = \frac{1}{2} \left(\frac{\overrightarrow{f_{l,t}} - \overrightarrow{ r_{i,j}}}{\|f_{l,t} - r_{i,j}\|} 
    + \frac{\overrightarrow{s} - \overrightarrow{ r_{i,j}}}{\|s - r_{i,j}\|} \right),
    \label{eq:normal}
\end{equation}
\begin{equation}
\label{eq:theta_cal}
\theta_{i,j,t} = \arccos\!\left( \overrightarrow{n_{i,j,t}} \cdot \hat{z} \right),
\end{equation}
\begin{equation}
\label{eq:phi_cal}
\phi_{i,j,t} = \operatorname{atan2}\big( \overrightarrow{n_{i,j,t}} \cdot \hat{y}, \overrightarrow{n_{i,j,t}} \cdot \hat{x}\big),
\end{equation}
where $r_{i,j} \in \mathbb{R}^3$ is the position of tile $(i,j)$, $s \in \mathbb{R}^3$ denotes the access point location, and $\hat{x}, \hat{y}, \hat{z}$ are unit vectors defining the coordinate frame.

Physical constraints impose:
\begin{equation}
\theta_{i,j,t} \in [\theta_{\min}, \theta_{\max}], \quad
\phi_{i,j,t} \in [\phi_{\min}, \phi_{\max}], \quad \forall i,j.
\end{equation}
\subsubsection{Reward Function}
The system reward is designed to maximize aggregate received signal strength:
\begin{equation}
R(s_t, a_t) = \frac{1}{K} \sum_{k=1}^K P_{r,k}(s_t, a_t),
\end{equation}
where $P_{r,k}(s_t,a_t)$ is the received signal strength indicator (RSSI) for user $k$.

\subsubsection{Optimization Objective}

The multi-agent system seeks a joint policy
\[
\pi = \{\pi_1, \pi_2, \dots, \pi_L\},
\]
that maximizes the expected discounted cumulative reward:
\begin{equation}
J(\pi) = \mathbb{E}_{\tau \sim \pi}\left[\sum_{t=0}^\infty \gamma^t R(s_t, a_t)\right],
\end{equation}
where $\tau$ represents a trajectory generated by following policy $\pi$.

This formulation decomposes complex system-wide optimization into agent-specific sub-problems while maintaining coordination via the shared reward signal and global state information. By operating at the spatial geometry level, it eliminates the need for explicit channel state estimation, significantly reducing computational complexity compared to conventional RIS approaches.

\subsection{Physical System Model}

\begin{figure}[!htp]
    \centerline{\includegraphics[width=0.8\linewidth]{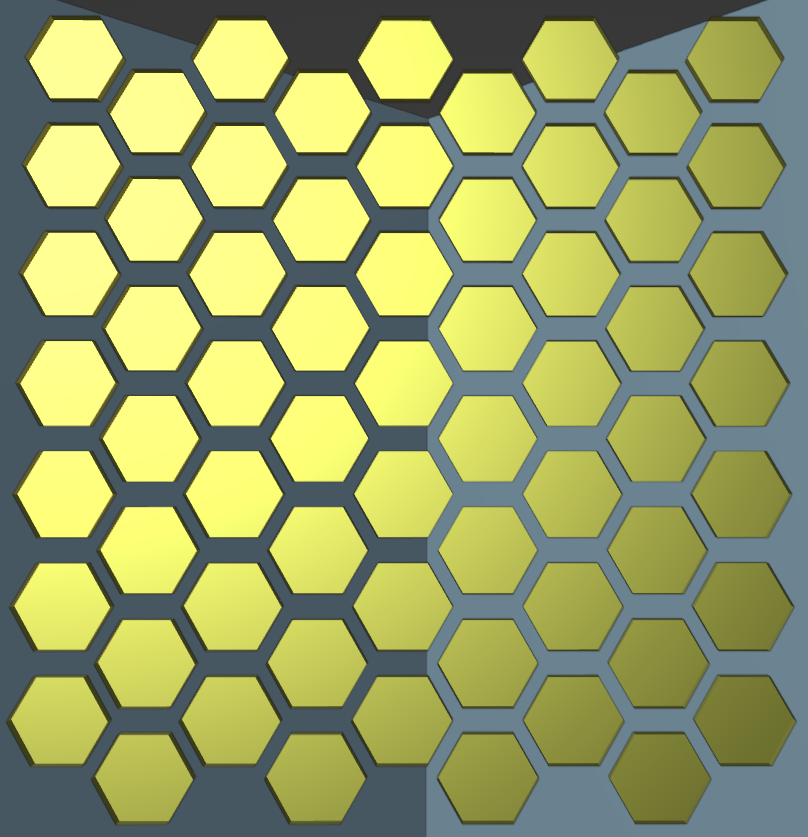}}
    \caption{Reflector with hexagonal elements.\label{Figure:reflector}}
\end{figure}

Our surface is an $N_r \times N_c$ array of hexagonal metallic tiles (Fig. \ref{Figure:reflector}), each adjustable in elevation and azimuth to steer mmWave beams. This physical reconfigurability enables precise control over electromagnetic wavefront manipulation without requiring complex RF circuits or electronic phase shifters.

\subsubsection{Focal Point Control Mechanism}
The core innovation of our approach lies in the focal point control paradigm, which abstracts the complex individual tile control problem into a geometrical framework. Each agent $l$ controls a movable focal point $f_{l,t}$ that determines the reflection characteristics of all tiles assigned to that agent. The focal point position evolves according to:
\begin{equation}
    f_{l,t+1} = f_{l,t} + a_{l,t},
\end{equation}
subject to the physical constraint:
\begin{equation}
    f_{l,t} \in F_l = \{f \in \mathbb{R}^3 : f_l^{\min} \leq f \leq f_l^{\max}\},
\label{eq:focal_constraints}
\end{equation}
where $f_l^{\min}$ and $f_l^{\max}$ define the allowable focal point region for agent $l$.

\subsubsection{Physical Constraint Enforcement}
The mechanical limitations of the servo-actuated reflector elements impose bounds on achievable orientations:
\begin{equation}
    \theta_{\min} \leq \theta_{i,j,t} \leq \theta_{\max}, \quad \forall (i,j),
\label{eq:theta_constraints}
\end{equation}
\begin{equation}
    \phi_{\min} \leq \phi_{i,j,t} \leq \phi_{\max}, \quad \forall (i,j),
\label{eq:phi_constraints}
\end{equation}
Typical values for practical implementations are $\theta_{\min} = -\pi/6$, $\theta_{\max} = \pi/6$, $\phi_{\min} = -\pi/6$, and $\phi_{\max} = \pi/6$, representing $\pm30^\circ$ rotation capability in both axes.

\subsubsection{Agent-to-Reflector Element Assignment}
The reflector array is partitioned into $L$ disjoint segments, where each segment $S_l$ is controlled by agent $l$:
\begin{equation}
S = S_1 \cup S_2 \cup ... \cup S_L,\quad S_{i'} \cap S_{j'} = \emptyset, \forall i' \neq j'
\end{equation}

The assignment function
\[
\Psi: \{1, 2, ..., N_r\} \times \{1, 2, ..., N_c\} \rightarrow \{1, 2, ..., L\},
\]
maps each reflector element to its controlling agent:
\begin{equation}
\Psi(i,j) = l \iff (i,j) \in S_l
\end{equation}

\subsubsection{Signal Propagation Model}
The received signal strength at user $k$ is determined by the aggregate contribution from all multipath components, including NLOS transmission and reflector-assisted paths. The total received power is {expressed as:}
\begin{equation}
P_{r,k}(t) = P_t \left| \sum_{(i,j)} h_{r,k}^{(i,j)}(t) + h_{\text{wall}}(t) \right|^2,
\end{equation}
where $P_t$ represents the transmitted power, $h_{\text{wall}}(t)$ and $h_{r,k}^{(i,j)}(t)$ represent the cascaded channel coefficient through surrounding walls and reflector element $(i,j)$ at time $t$, respectively.

\subsubsection{Computational Complexity Reduction}

The focal point control approach provides substantial dimensionality reduction, with the control space reduced from $2N_rN_c$ individual angular parameters to $3L$ focal point coordinates, where L represents the number of agents and typically $L \ll N_rN_c$. The complexity reduction factor is quantified as:
\begin{equation}
R_{\text{complexity}} = \frac{2N_rN_c}{3L},
\end{equation}
For a deployment with $N_{\text{total}} = N_rN_c = 72$ tiles serving $K = 3$ users with  $L = 3$ agents, the complexity reduction factor reaches $R_{\text{complexity}} \approx 16$, representing substantial computational advantage. Physical constraints are automatically satisfied through geometric relationships, eliminating explicit constraint enforcement during policy execution, while providing intuitive interpretability for human operators observing focal point positions relative to user locations.

\subsection{Optimization Problem Formulation}

The intelligent reflecting surface control problem can be formulated as a constrained multi-objective optimization challenge that seeks to maximize aggregate system performance while respecting physical limitations and practical deployment constraints. Our formulation operates at the spatial geometry level through focal point control, significantly reducing computational complexity while maintaining effective beam steering capabilities.

\subsubsection{Primary Optimization Objective}

The reflector tile control constitutes a constrained multi-objective optimization challenge maximizing aggregate system performance while respecting physical limitations. The constraints of the system follows (\ref{eq:focal_constraints}), (\ref{eq:theta_constraints}), and (\ref{eq:phi_constraints}). The optimization operates at the spatial geometry level through focal point control, significantly reducing computational complexity while maintaining effective beam steering capabilities. Therefore, the complete constrained optimization problem becomes:

\begin{maxi!}|s|
    {f_1,...,f_L}{\sum_{k=1}^K P_{r,k}(f_1,...,f_L, u_k) \label{eq:problem_objective}}
    {}{}
    \addConstraint{f_l \in F_l, \ \forall l \in \{1, 2, ..., L\} \label{eq:p1_constraint1}}
    \addConstraint{\theta_{i,j}(f_l) \in [\theta_{\min}, \theta_{\max}], \ \forall(i,j) \in S_l, \forall l \label{eq:p1_constraint2}}
    \addConstraint{\phi_{i,j}(f_l) \in [\phi_{\min}, \phi_{\max}], \ \forall(i,j) \in S_l, \forall l \label{eq:p1_constraint3}},
\end{maxi!}
where $P_{r,k}(f_1,...,f_L, u_k)$ denotes the received power at user $k$ as a function of all agent focal point positions ($(f_1,...,f_L$) and user location $u_k$. $\theta_{i,j}(f_l)$ and $\phi_{i,j}(f_l)$ are computed according to (\ref{eq:theta_cal}) and (\ref{eq:phi_cal}), ensuring all resulting tile angles remain within servo motor capabilities.

{It is important to note that the optimization constraints in (\ref{eq:p1_constraint1})-(\ref{eq:p1_constraint3}) are enforced as hard constraints within the simulation environment. The focal point positions are clipped to the region $F_l$ at every time step, and the tile orientations derived from these focal points are mechanically limited to the servo's operational range. Consequently, the agent's action space is naturally bounded to the feasible set, removing the need for penalty terms or Lagrange multipliers in the reward function that are typically required for Constrained MDPs.}

\subsubsection{Dynamic Optimization Considerations}
The optimization problem extends to dynamic scenarios where user positions evolve over time according to mobility patterns. The time-varying optimization becomes:
\begin{equation}
    \max_{f_1(t),...,f_L(t)} \mathbb{E}\left[\sum_{t=0}^T \sum_{k=1}^K P_{r,k}(f_1(t),...,f_L(t), u_k(t))\right]
\end{equation}
subject to the same spatial constraints (\ref{eq:p1_constraint1}-\ref{eq:p1_constraint3}) holding at each time instant $t$, plus additional temporal constraints ensuring smooth focal point transitions:
\begin{equation}
    \|f_l(t+1) - f_l(t)\|_2 \leq v_{\max}\Delta t
\end{equation}
where $v_{\max}$ represents the maximum focal point velocity and $\Delta t$ is the discrete time step.

\subsubsection{Solution Approach Justification}

{
The formulated optimization problem exhibits characteristics making traditional techniques computationally prohibitive: non-convexity due to multipath propagation effects, high dimensionality despite reduction, partial observability in practical deployments, and dynamic environments with time-varying landscapes. These characteristics motivate MARL adoption, effectively handling non-convex, high-dimensional, partially observable optimization problems while adapting to dynamic conditions through continuous learning.

Focal point control also facilitates efficient handling of LOS and NLOS scenarios through adaptive positioning strategies. When direct paths between the access point and users are available, agents can position their focal points to create concentrated signal enhancement in user locations. In NLOS scenarios, agents can strategically position focal points to exploit wall reflections or other environmental features, creating indirect propagation paths that maintain connectivity despite obstacles. This flexibility enables the system to automatically adapt to diverse propagation conditions without requiring explicit environment-specific programming.
}

\subsection{Simulation Framework and Validation}

\begin{figure}[!t]
    \centering
    \includegraphics[width=1\linewidth]{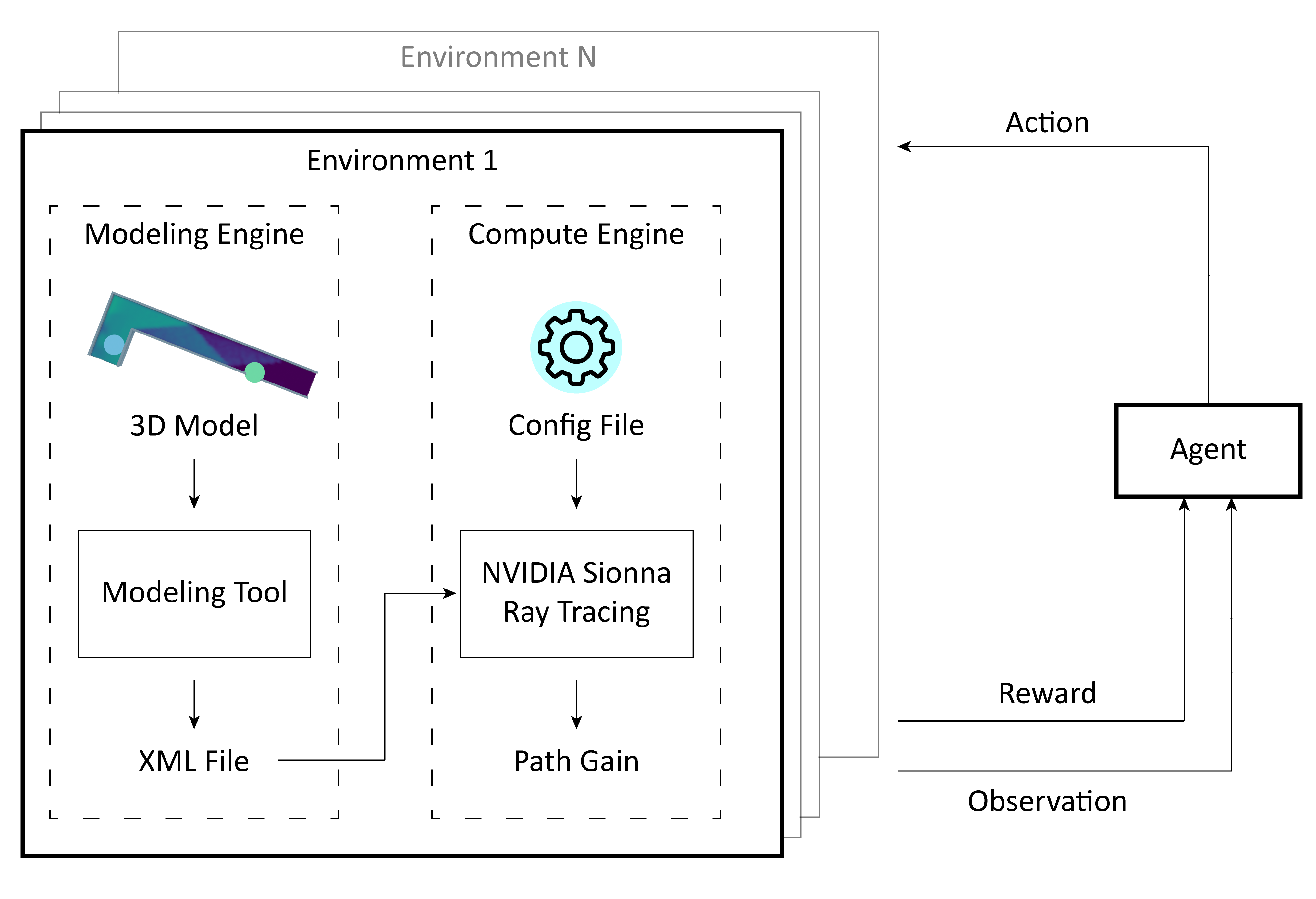}
    \caption{Workflow of the Deep Reinforcement Learning and the integrated environment with Sionna and Blender.}
    \label{Figure:workflow}
\end{figure}

The validation of the proposed multi-agent focal point control methodology requires a comprehensive simulation framework capable of accurately modeling electromagnetic propagation, mechanical constraints, and dynamic environmental conditions. Our simulation environment integrates state-of-the-art ray-tracing techniques with realistic material properties and precise geometric modeling to ensure that the learned policies translate effectively to practical deployment scenarios. The framework (shown in Fig. \ref{Figure:workflow}) combines NVIDIA Sionna's deterministic propagation engine \cite{sionna} with Blender's 3D modeling capabilities \cite{blender}, creating a seamless pipeline for automated wireless signal calculations and reflector configuration optimization.

\subsubsection{Ray-Tracing Propagation Model}

The simulation employs deterministic 3D ray-tracing methodology to model mmWave signal propagation, an approach well-suited for high-frequency wireless systems where dominant propagation mechanisms closely resemble optical behavior. The ray-tracing engine simulates electromagnetic wave interactions through multiple environmental bounces, capturing reflection, diffraction, and scattering effects that significantly influence signal strength in complex indoor environments. NVIDIA Sionna's propagation engine integrates detailed 3D geometry with material-specific electromagnetic properties, enabling accurate modeling of cascaded reflections and transmission losses that characterize realistic deployment scenarios.

The ray-tracing approach provides several advantages over simplified statistical channel models. First, it captures environment-specific geometric effects with high fidelity \cite{a10686755}, accounting for precise reflector element positions and orientations. Second, it naturally incorporates multipath propagation effects arising from environmental obstacles and reflecting surfaces \cite{a10416965}. Third, it enables dynamic channel modeling with time-varying multipath delays and Doppler shifts.

\subsubsection{Material Characterization}

To ensure simulation accuracy and practical relevance, all environmental materials are characterized according to International Telecommunications Union (ITU) recommendations for electromagnetic properties. The material parameter specification follows ITU-R Recommendation P.2040 \cite{rec_p2040}, which provides empirically-derived frequency-dependent constants for relative permittivity and conductivity calculations. The mathematical models for material properties are expressed as:
\begin{equation}
    \eta'(f) = a f^b,
\end{equation}
\begin{equation}
    \sigma_c(f) = c f^d,
\end{equation}
where $\eta'$ represents the dimensionless relative permittivity, $\sigma_c$ denotes conductivity in S/m, $f$ is the operating frequency in GHz, and the constants $a$, $b$, $c$, $d$ are material-specific parameters derived from extensive measurement campaigns\cite{rec_p2040}.

\subsubsection{Automated Simulation Pipeline}

The simulation framework implements an automated pipeline that seamlessly integrates 3D modeling, electromagnetic simulation, and reinforcement learning training. The pipeline begins with environmental setup in Blender, where structural elements including walls, floors, ceilings, obstacles, and the adjustable reflector array are positioned according to experimental specifications. Each reflector element is modeled as an independently controllable hexagonal tile with mechanical rotation constraints matching practical servo motor capabilities.

The focal point control system interfaces with Blender through a RESTFUL API that automatically calculates required tile orientations based on agent-specified focal point positions. When an agent updates its focal point coordinates, the system computes the optimal surface normal for each associated reflecting element using equation (35), then derives the corresponding elevation and azimuth angles through geometric transformations. The configured 3D model is exported in XML format compatible with Sionna's ray-tracing engine, ensuring consistent electromagnetic simulation across different reflector configurations.

\subsubsection{Dynamic Environmental Modeling}

The simulation framework accommodates dynamic scenarios through time-varying user positions and channel evolution. User mobility is modeled through stochastic movement patterns that reflect realistic indoor navigation behaviors. The system updates user positions at regular intervals, triggering automatic recalculation of optimal focal point positions and corresponding reflector element orientations. This dynamic capability enables evaluation of system adaptation performance and policy robustness under realistic operational conditions.

Environmental obstacles are positioned to create challenging non-line-of-sight scenarios that test the system's ability to exploit indirect propagation paths. A representative obstacle configuration includes elliptical obstructions positioned to block direct paths between the access point and user equipment, forcing the system to rely primarily on reflector-assisted signal enhancement. The obstacle dimensions and material properties are selected to accurately represent typical impediments that affect mmWave propagation.

\subsubsection{Integration with Learning Algorithms}

The simulation framework provides standardized interfaces for reinforcement learning algorithm integration, enabling direct policy evaluation and training without manual intervention. The system automatically generates state observations based on user locations and current focal point positions, computes reward signals from simulated RSSI measurements, and applies policy-generated actions to update reflector configurations. This seamless integration ensures that the learning process operates on realistic channel conditions while maintaining consistent environmental modeling throughout training and evaluation phases.

The framework supports both centralized training scenarios where global system state is available, and decentralized execution modes where agents operate based solely on local observations. This dual-mode capability enables validation of the CTDE paradigm under realistic information constraints while ensuring that learned policies remain practically deployable. Performance metrics including received signal strength, adaptation time, and configuration feasibility are automatically logged to enable comprehensive analysis of system behavior across diverse operational conditions.

\section{MULTI-AGENT REINFORCEMENT LEARNING WITH CENTRALIZED TRAINING WITH DECENTRALIZED EXECUTION}
\label{sec:marl_ctde}

\subsection{Centralized Training with Decentralized Execution Framework}

{The CTDE paradigm addresses fundamental challenges in multi-agent learning by leveraging global information during training while maintaining practical deployment constraints. The CTDE approach resolves the non-stationarity problem inherent in multi-agent environments, where each agent's environment appears non-stationary due to simultaneous learning by other agents.

During centralized training, a global critic network $V^\pi(s_{\mathrm{global}})$ accesses the complete system state $s_{\mathrm{global}}$, enabling accurate value estimation of complex dynamics. The critic computes the expected return under the joint policy $\pi$ as:}
\begin{equation}
    V^\pi(s_{\mathrm{global}}) = \mathbb{E}_{\pi}\Big[G_t\;|\;S_{\mathrm{global},t}=s_{\mathrm{global}}\Big],
\end{equation}
{where $G_t$ denotes the cumulative discounted return from time $t$ onward. The global critic thus relates joint actions to system-wide rewards for superior advantage estimation. The advantage function, leveraging global state data, guides effective policy updates:}
\begin{equation}
    \text{Adv}(s_{\mathrm{global}}, a_l) = Q^\pi(s_{\mathrm{global}}, a_l) - V^\pi(s_{\mathrm{global}}),
\end{equation}
{where $Q^\pi(s_{\mathrm{global}}, a_l)$ is the expected return for agent $l$ taking action $a_l$ in state $s_{\mathrm{global}}$.

In decentralized execution, each agent operates independently using only its local observation $o_l$, with policies $\pi_{\theta_l}(a_l\,|\,o_l)$ mapping observations to actions. No inter-agent communication is required, maintaining scalability and feasibility for practical deployments while preserving the coordination learned during centralized training.

The use of global state information in advantage computation provides several theoretical advantages. First, the non-stationarity mitigation occurs because the centralized critic observes the complete system state, providing stable value estimates despite simultaneous learning by multiple agents. Second, enhanced credit assignment results from global state information enabling better attribution of rewards to individual agent actions. The theoretical convergence of CTDE relies on the assumption that the environment becomes stationary from the perspective of the global critic, even though individual agents experience non-stationary local environments.
}

\subsection{Multi-Agent Proximal Policy Optimization}

Multi-Agent Proximal Policy Optimization (MAPPO) extends the successful single-agent Proximal Policy Optimization (PPO) algorithm \cite{schulman2017proximal} to multi-agent environments while addressing unique challenges of coordination and non-stationarity. The algorithm maintains the clipped surrogate objective that constrains policy updates to prevent destructive changes, enhanced with global state information from the CTDE framework \cite{yu2022surprising}.

For each agent $l$, the clipped surrogate objective becomes:

\begin{equation}
\begin{aligned}
L^{CLIP}(\psi_l) = \mathbb{E}_t \Big[ \min\Big( r_t^l(\psi_l) \text{Adv}(s_{global,t}, a_t^l), \\
\text{clip}(r_t^l(\psi_l), 1{-}\epsilon, 1{+}\epsilon) \text{Adv}(s_{global,t}, a_t^l) \Big) \Big],
\end{aligned}
\end{equation}
where the probability ratio $r_t^l(\psi_l) = \frac{\pi_{\psi_l}(a_t^l|o_t^l)}{\pi_{\psi_{old,l}}(a_t^l|o_t^l)}$ measures the change in policy $\psi_l$ between updates, and $\text{Adv}(s_{global,t}, a_t^l)$ represents the advantage function computed using global state information. The operator "clip" represents a clipping function that constrains the value of $r_t^l(\psi_l)$ within the interval $[1{-}\epsilon, 1{+}\epsilon]$. The clipping parameter $\epsilon$ (typically 0.1-0.2) constrains policy changes to maintain training stability.

The value function loss for the centralized critic is:

\begin{equation}
L^{VF}(\xi) = \mathbb{E}_t \left[ \left( V_{\xi}(s_{global,t}) - V_{target,t} \right)^2 \right],
\end{equation}
where $\xi$ is the neural network parameters for value function $V_{\xi}(s_{global,t})$. $V_{target,t}$ represents the target value computed using trajectory data $(S_t, A_t, R_t)$ representing state, action, and reward at time step $t$.

The complete MAPPO objective combines policy and value function losses:

\begin{equation}
L(\psi_l, \xi) = L^{CLIP}(\psi_l) - c_1 L^{VF}(\xi) + c_2 H[\pi_{\psi_l}](o_t^l),
\end{equation}
where $H[\pi_{\psi_l}](o_t^l)$ represents the entropy bonus to encourage exploration, and $c_1$, $c_2$ are weighting coefficients for value function loss and entropy regularization, respectively.

The MAPPO algorithm demonstrates several stability properties crucial for practical reflector control deployment. The clipped objective function prevents large policy updates that could destabilize the coordination patterns learned by interacting agents. The entropy regularization term maintains exploration capabilities throughout training, preventing premature convergence to suboptimal coordination strategies. The centralized critic provides stable value estimates despite the non-stationarity induced by simultaneous agent learning, enabling reliable advantage estimation for policy updates.

\subsection{{Stability and Convergence Characteristics}}

{While rigorous convergence proofs for multi-agent reinforcement learning remain an open research challenge due to non-stationarity of the environment, MAPPO with CTDE employs several mechanisms to enhance stability. The algorithm relies on the theoretical foundations of single-agent PPO \cite{a10663867}, specifically the clipped surrogate objective:}
\begin{equation}
\mathbb{E}[\eta(\pi_{new})] \geq \mathbb{E}[\eta(\pi_{old})] - C \cdot \mathbb{E}[D_{KL}(\pi_{old}, \pi_{new})],
\end{equation}
where $\eta(\pi)$ represents the expected return under policy $\pi$, and $C$ is a problem-dependent constant.

{In single-agent settings, this objective creates a lower bound on policy improvement \cite{wang2020truly}. In our multi-agent framework, the CTDE paradigm acts to satisfy the stationarity assumptions required for this bound to hold approximately. By feeding the global state $s_{global}$ into the centralized critic, the value function $V(s_{global})$ accounts for the joint actions of all agents, stabilizing the advantage estimates $\hat{A}_t$.}

{Although \cite{yu2022surprising} primarily provides empirical validation rather than theoretical proofs for MAPPO, it demonstrates that this PPO-based approach achieves state-of-the-art performance in cooperative tasks by effectively managing the trust region of policy updates. Our empirical results (Fig. 5) align with these findings, showing low variance and consistent reward growth, suggesting that the centralized critic successfully mitigates the non-stationarity that typically hinders multi-agent convergence.}

\section{RESULTS AND DISCUSSION}
\label{sec:results_and_discussion}

{
We next quantify MARL-driven beam focusing in a 60 GHz downlink under NLOS conditions, using ray-tracing to compare RSSI across five control schemes
}

\subsection{{Experimental Setup}}

\begin{figure}[!t]
    \centering
    \captionsetup{justification=centering}
    \includegraphics[width=0.6\linewidth]{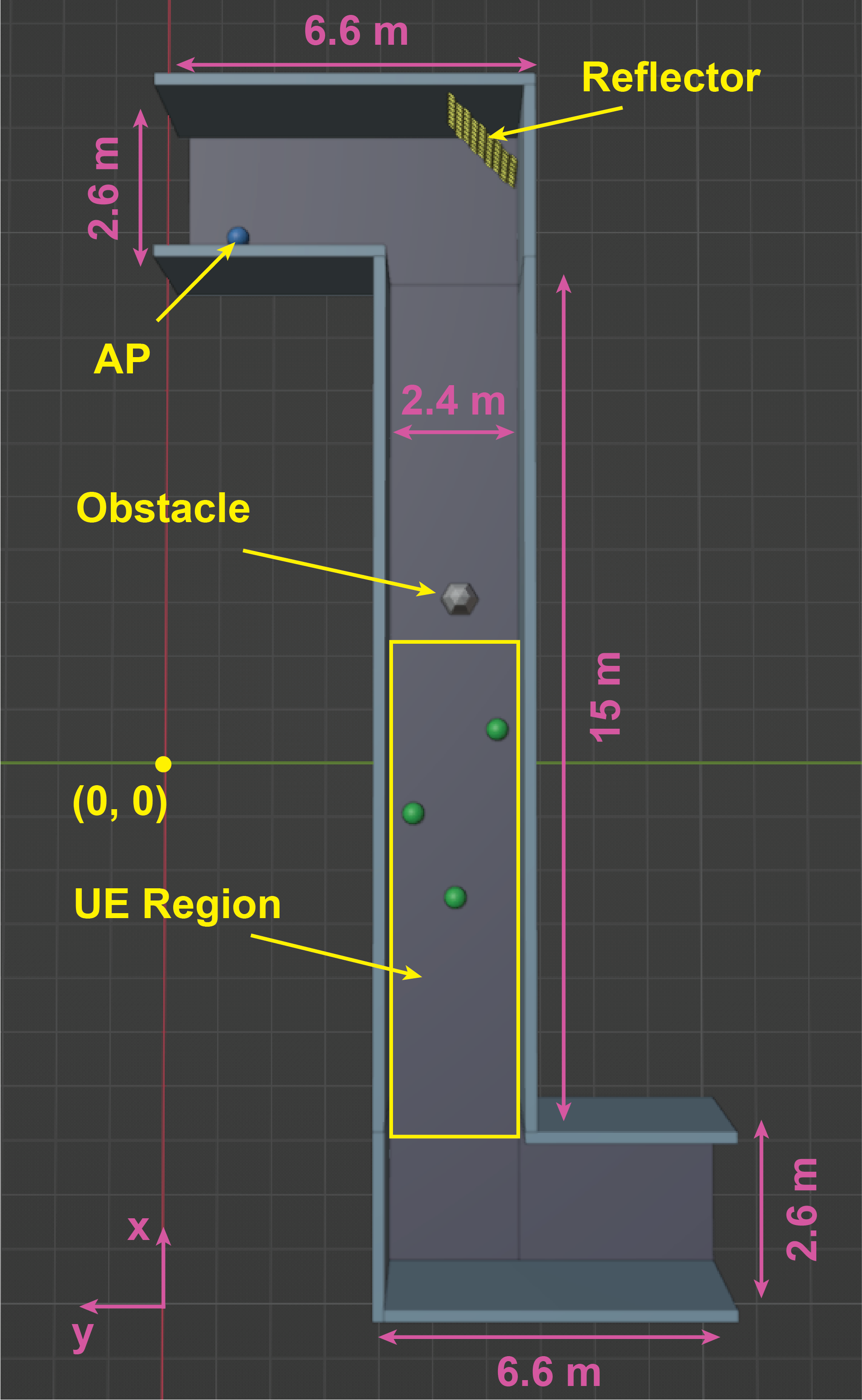}
    \caption{{Experimental setup of an L-shaped hallway for simulation. The AP is depicted in blue, while the users are shown in pink. The obstruction is represented in red.}}
    \label{Figure:experimental_setup}
\end{figure}

{
Experiments are conducted in an L-shaped hallway to analyze wireless signal propagation in the presence of obstacles. The experimental layout is illustrated in Fig.~\ref{Figure:experimental_setup}, which also indicates the dimensions of the environment. The vertical segment of the hallway extends approximately $15~\mathrm{m}$, while the total length from end to end is about $20~\mathrm{m}$. Both hallway ends are open to emulate realistic indoor deployment scenarios better.

Within the setup, a single access point (AP) equipped with 4 antennas is positioned at coordinates $(9.5,\,-1.5)$ (where the first value denotes the $x$-coordinate and the second the $y$-coordinate) at a height of $2.5~\mathrm{m}$ above the floor. The total transmit power is set to 5.0 dBm. A solid oval obstacle, representing an object with a height of $1.8~\mathrm{m}$, is placed at $(3.0,\,-5.5)$. The reflector, consisting of $72$ hexagonal tiles, is installed at $(11.2, -5.7)$ as shown in Fig.~\ref{Figure:experimental_setup}. Three user equipments (UEs), each equipped with a single antenna and depicted in green, are distributed within the designated UE region. The locations of the UEs are either fixed or uniformly randomized within the region spanning from $(-6.0,\,-6.2)$ to $(2.0,\,-4.25)$. Further details regarding user placement are provided in subsequent sections.

The 3D model incorporates a variety of materials to closely mimic real-world building structures. The walls are constructed from plasterboard, the ceiling from board material, and the floor from concrete, while the obstacle is made of wood. The reflectivity and conductivity parameters for all materials are referenced from ITU Recommendation~\cite{rec_p2040}.

To reflect practical deployment conditions, the initial orientation of the reflector tiles is uniformly randomized within the physical range of $[-\nicefrac{\pi}{6},\,\nicefrac{\pi}{6}]$ at the start of each experimental episode. The procedure is designed to simulate the variability in reflector configurations that may occur prior to optimizing signal reflection toward intended users.
}

\subsection{Hyperparameter Configuration}

\begin{table}[ht!]
\caption{MAPPO Hyperparameters for Intelligent Reflecting Surface Control.}
\centering
\begin{tabular}{l|l}
    \hline
    Parameter & Value \\
    \hline
    Number of neurons (hidden layers) & 256 \\
    Optimizer & Adam \\
    Learning rate & $2.0 \times 10^{-4}$ \\
    Learning rate schedule & Constant \\
    Discount factor ($\gamma$) & 0.985 \\
    Replay buffer size & 1000 \\
    Minibatch size & 200 \\
    Activation function & ReLU \\
    PPO clipping parameter ($\epsilon_{CLIP}$) & 0.2 \\
    GAE parameter ($\lambda_{GAE}$) & 0.9 \\
    Value function coefficient ($c_1$) & 0.5 \\
    Entropy coefficient ($c_2$) & $1.0 \times 10^{-4}$ \\
    \hline
\end{tabular}
\label{table:hyperparameters}
\end{table}

{
The MAPPO implementation for intelligent reflecting surface control requires careful hyperparameter selection to ensure stable training dynamics and effective policy convergence in multi-agent environments. The configuration addresses unique multi-agent learning challenges including non-stationarity from simultaneously learning agents and coordination emergence across reflecting elements.

The hyperparameters used are outlined in Table \ref{table:hyperparameters}. The neural network architecture employs two fully connected hidden layers with 256 neurons each with ReLU activation function, providing sufficient representation capacity for complex spatial coordination. The Adam optimizer with constant learning rate of $2.0 \times 10^{-4}$ balances learning speed with training stability.

The discount factor $\gamma = 0.985$ balances long-term reward accumulation with maintaining sensitivity to immediate improvements. The replay buffer size of $1000$ transitions provides sufficient experience diversity for stable gradient estimation while maintaining relevance in non-stationary multi-agent environments. The minibatch size of $200$ samples enables robust gradient estimates while accommodating practical constraints of training multiple agent policies simultaneously.

The PPO clipping parameter $\epsilon_{CLIP} = 0.2$ constrains policy updates to prevent destabilization of learned coordination patterns. The generalized advantage estimation parameter $\lambda_{GAE} = 0.9$ balances bias-variance trade-offs in advantage estimation. The value function coefficient $c_1 = 0.5$ ensures the centralized critic develops accurate value estimates without overwhelming policy gradient signals, while the entropy coefficient $c_2 = 1.0 \times 10^{-4}$ provides modest exploration encouragement to prevent premature convergence to suboptimal coordination strategies.
}

\subsection{Training Performance Analysis and Algorithmic Comparison}

{
We conduct a comprehensive training analysis comparing three distinct approaches: multi-agent reinforcement learning with beam-focusing control (labeled as beam-focusing-ma), single-agent reinforcement learning with beam-focusing (labeled as beam-focusing-sa), and multi-agent learning with column-based control (labeled as column-based-ma). The beam-focusing approaches enable individual tile control with full degrees of freedom for both elevation and azimuth angles, allowing precise signal redirection toward specific points. In contrast, the column-based method introduces a cost-performance trade-off by restricting azimuth rotation to column-level control, where all elements within a column share the same azimuth angle while maintaining independent elevation control. This design significantly reduces servo requirements for physical implementation, requiring only one azimuth servo per column while maintaining element-level elevation control, thus reducing hardware complexity and cost at the expense of spatial degrees of freedom.
}

\begin{figure}[!t]
    \centering
    \captionsetup{justification=centering}
    \includegraphics[width=1\linewidth]{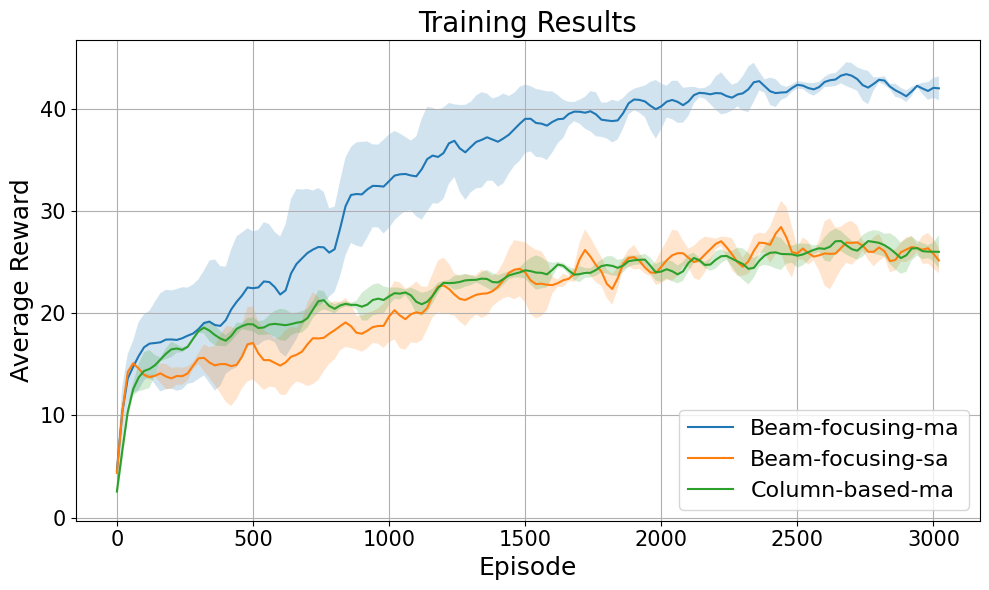}
    \caption{{Training performance comparison of MARL and single-agent approaches for reflecting element control. Multi-agent beam-focusing (beam-focusing-ma) achieves superior performance with a value of 42, outperforming single-agent beam-focusing (beam-focusing-sa) at a value of 27 and column-based multi-agent control (column-based-ma) at a value of 27. Solid lines represent mean training rewards with shaded regions indicating variance across multiple training runs.
}}
    \label{Figure:training}
\end{figure}

{
Fig. \ref{Figure:training} illustrates the training progression across 3000 episodes for the three methodologies. The beam-focusing-ma approach demonstrates superior learning performance, achieving rapid convergence to high reward values (approximately 42) and maintaining stable performance throughout training. The learning curve exhibits characteristic phases: initial rapid improvement during the first 1000 episodes, continued steady growth until episode 2000, followed by convergence behavior with marginal improvements. The multi-agent approach achieves faster initial learning and reaches higher performance levels more rapidly than the single-agent counterpart, with beam-focusing-ma demonstrating steeper learning curves and achieving reward values of approximately 40 cumulative reward during the first 1500 episodes, while beam-focusing-sa reaches only around 20 during the same period.

The performance hierarchy clearly emerges: beam-focusing-ma (42) $>$ column-based-ma (27) $\simeq$ beam-focusing-sa (27). The improved learning efficiency of the multi-agent approach can be attributed to the decomposition of the complex optimization problem into smaller, more manageable sub-tasks. Each agent focuses on serving a specific user, learning simplified coordination patterns rather than attempting to optimize the entire system's complex dynamics simultaneously. This task decomposition reduces the exploration space for individual agents while maintaining overall system coordination capability through the centralized training framework.

The column-based-ma approach shows more limited improvement potential, converging to lower reward values due to hardware-imposed constraints, yet demonstrates stable learning progression, validating that the MARL framework successfully adapts to restricted action spaces while maintaining learning stability. The training analysis provides valuable insights for practical deployment: the superior performance of beam-focusing-ma validates the proposed MARL framework effectiveness, with much greater convergence characteristics. The trade-off between beam-focusing and column-based approaches illustrates the performance cost of hardware simplification—while the column-based method achieves 35\% lower performance compared to full beam-focusing control, it offers substantial cost savings in servo requirements, making it attractive for cost-sensitive deployments where moderate performance reduction is acceptable.
}

\subsection{Comparative Performance Evaluation Against Baseline Methods}

\begin{figure}[!t]
  \subfloat[\centering No Reflector.\\(RSSI: $-108.2$ dBm)]{\includegraphics[width=0.3\linewidth]{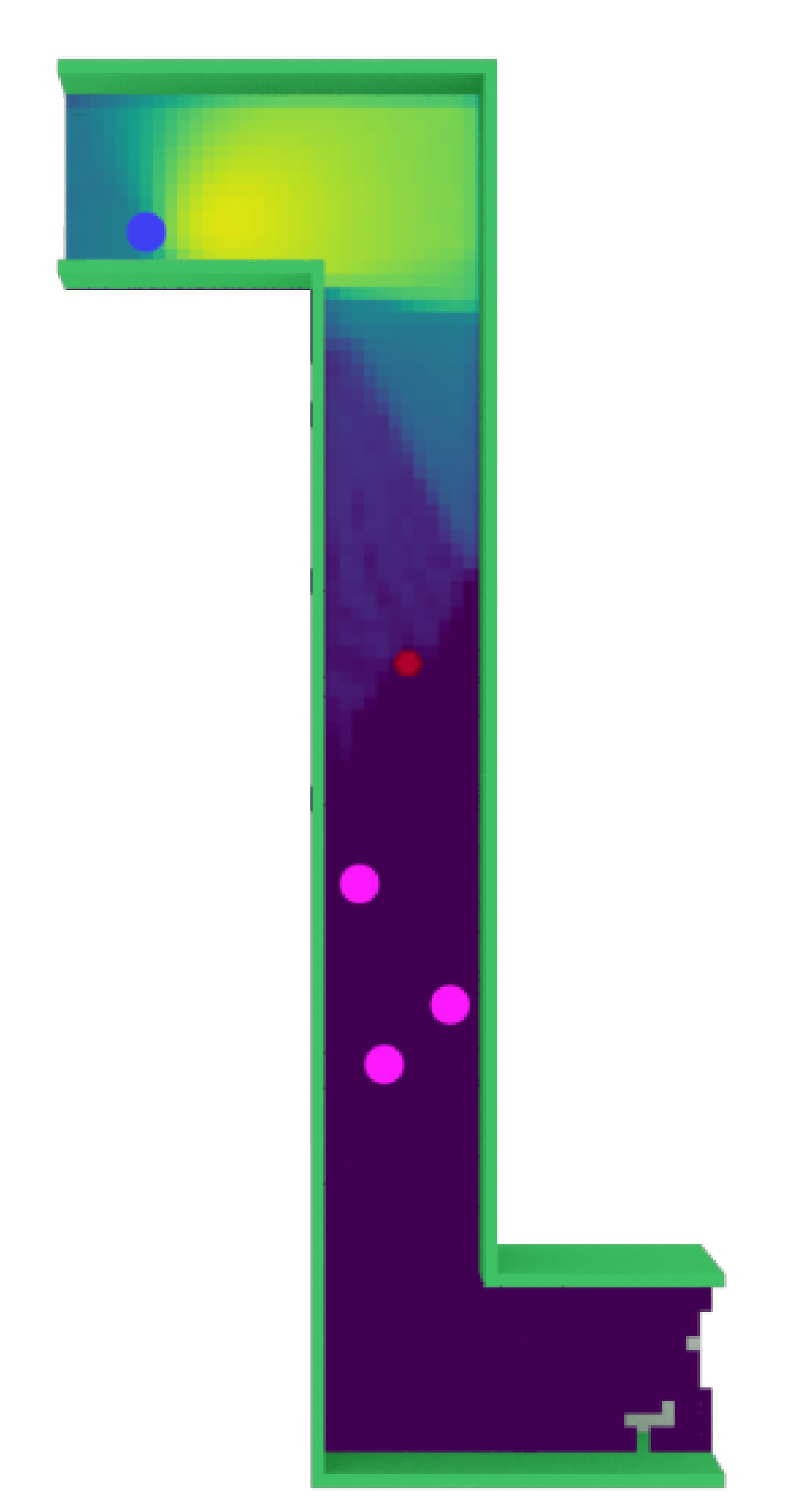}%
  \label{Figure:no_reflector}}
  \hfill
  \subfloat[\centering Flat Reflector.\\(RSSI: $-90.1$ dBm)]{\includegraphics[width=0.3\linewidth]{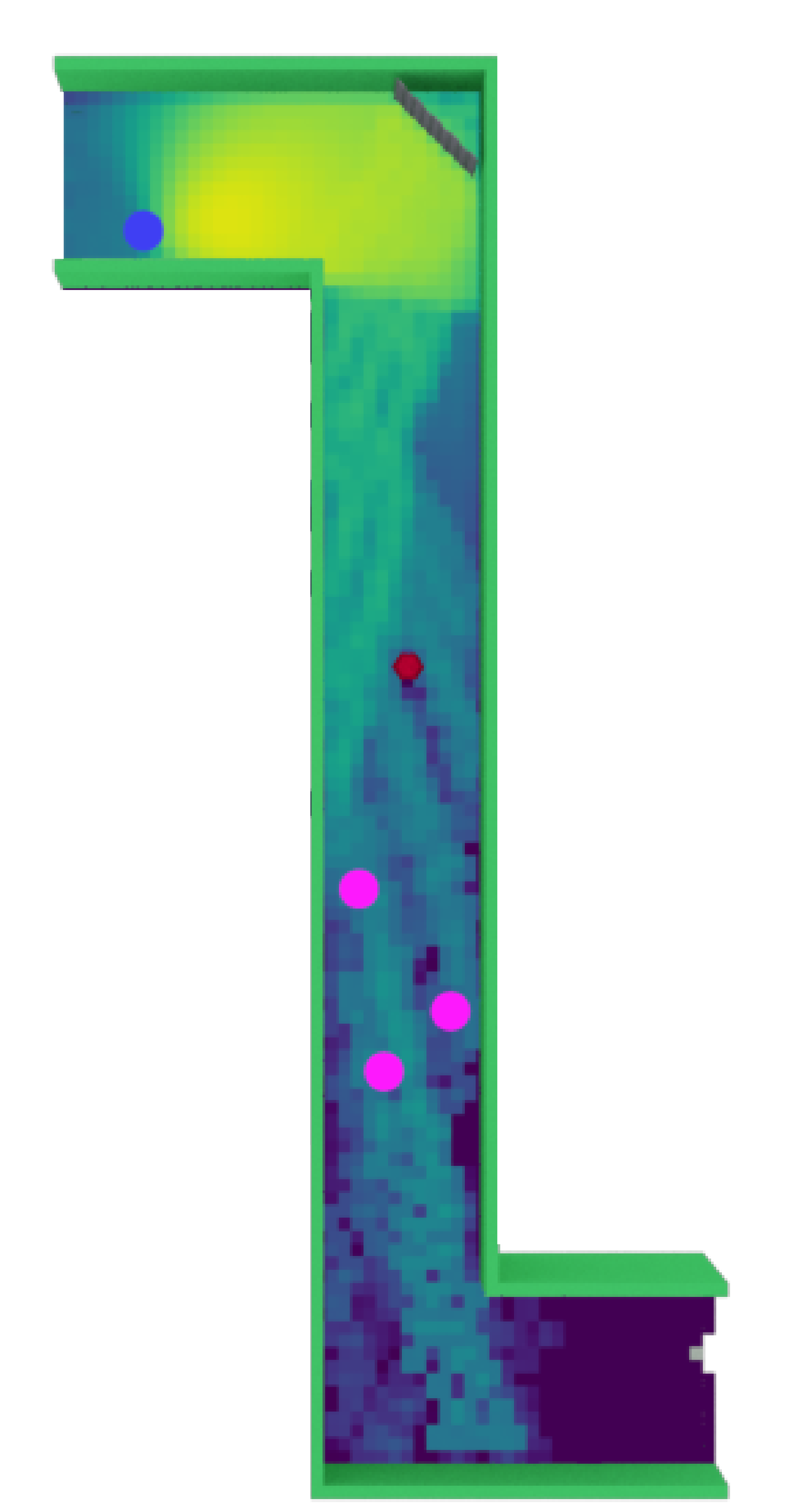}%
  \label{Figure:flat_reflector}}
  \hfill
  \subfloat[\centering SA Focusing. \\(RSSI: $-74.22$ dBm)]{\includegraphics[width=0.3\linewidth]{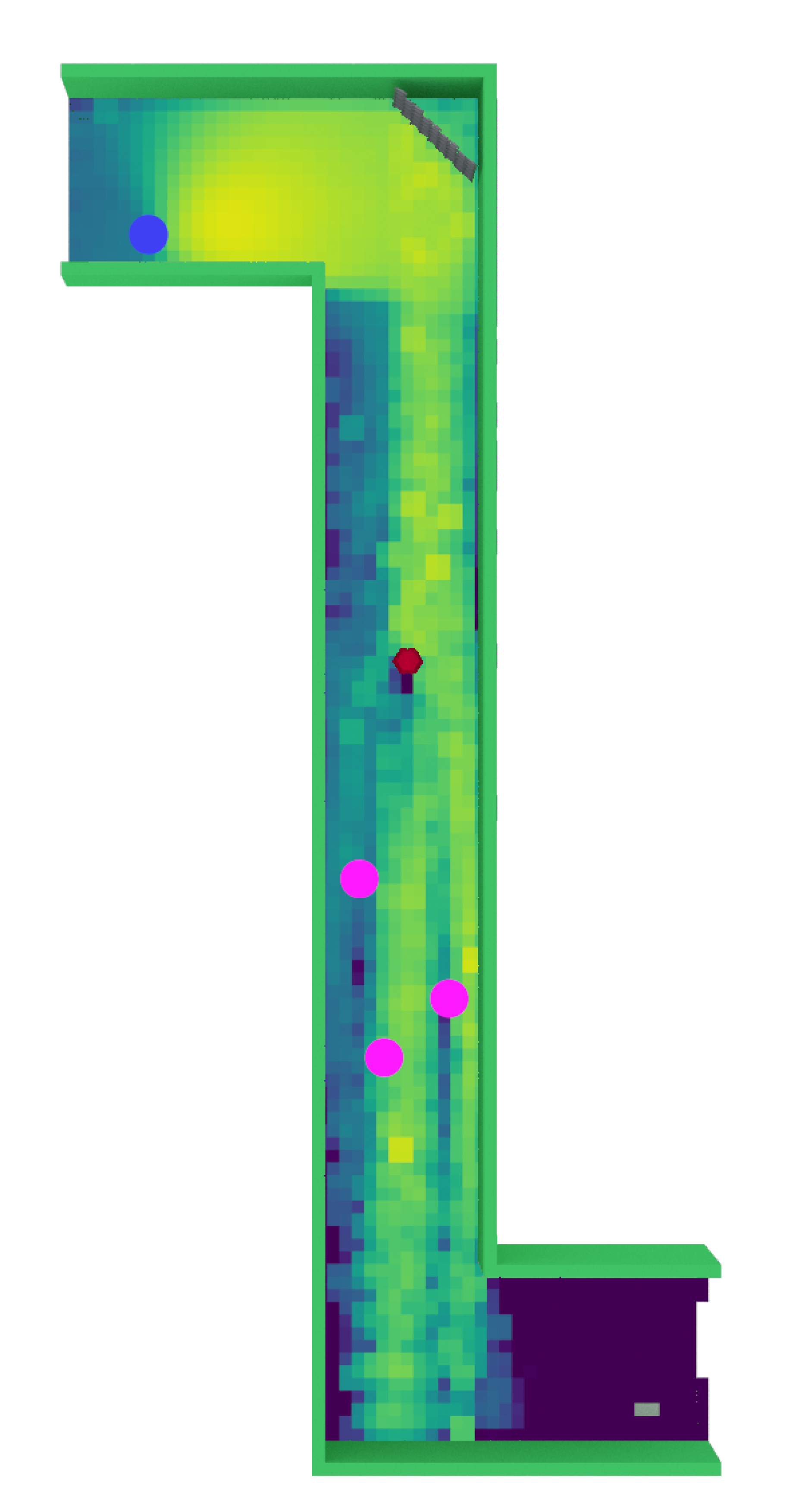}%
  \label{Figure:beamfocusing_sa}}
  \hfill
  \subfloat[\centering MA Col. \\(RSSI: $-72.71$ dBm)]{\includegraphics[width=0.3\linewidth]{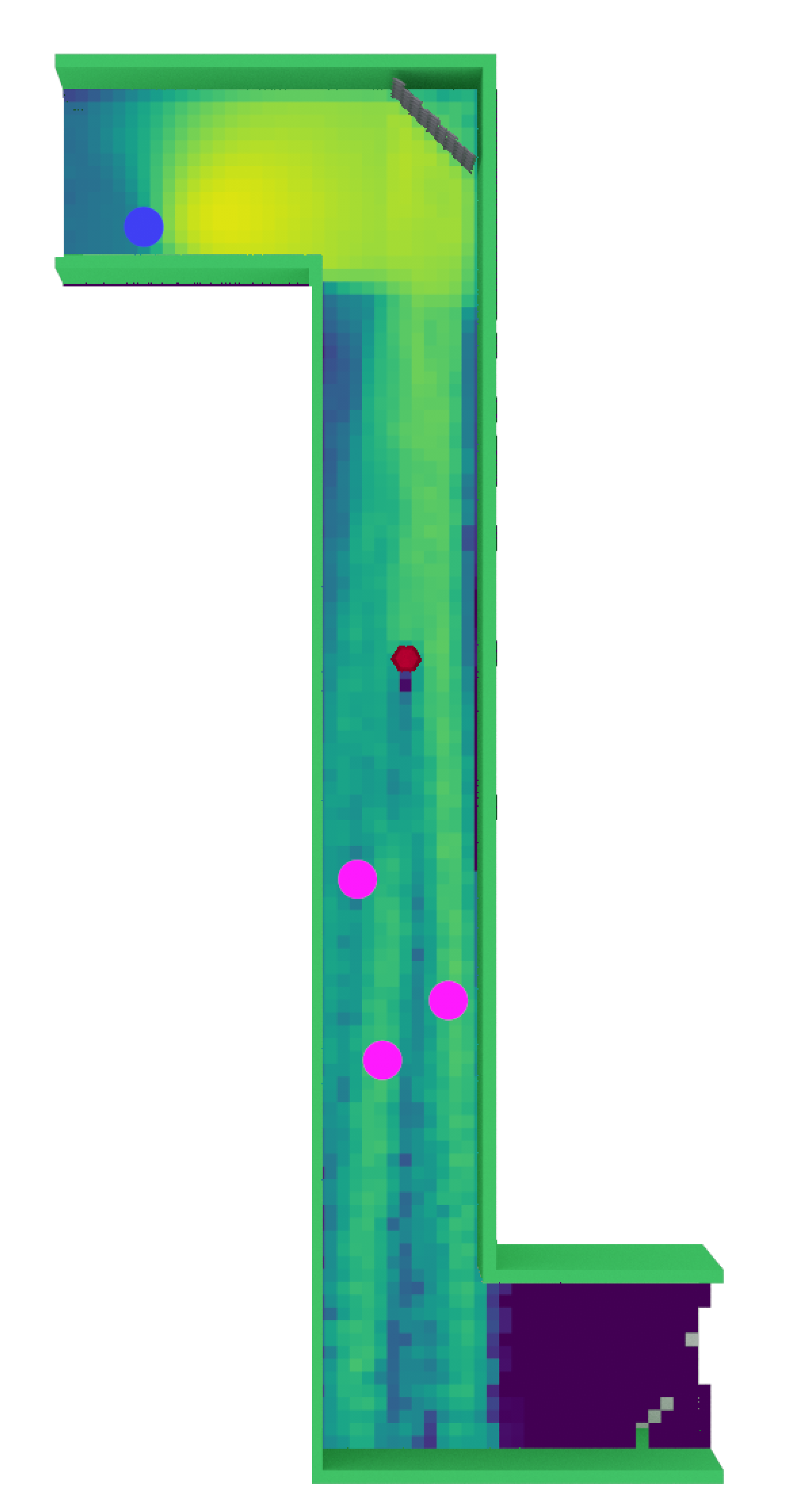}%
  \label{Figure:col_ma}}
  \hfill
  \subfloat[\centering MA Focusing. \\(RSSI: $-69.4$ dBm)]{\includegraphics[width=0.3\linewidth]{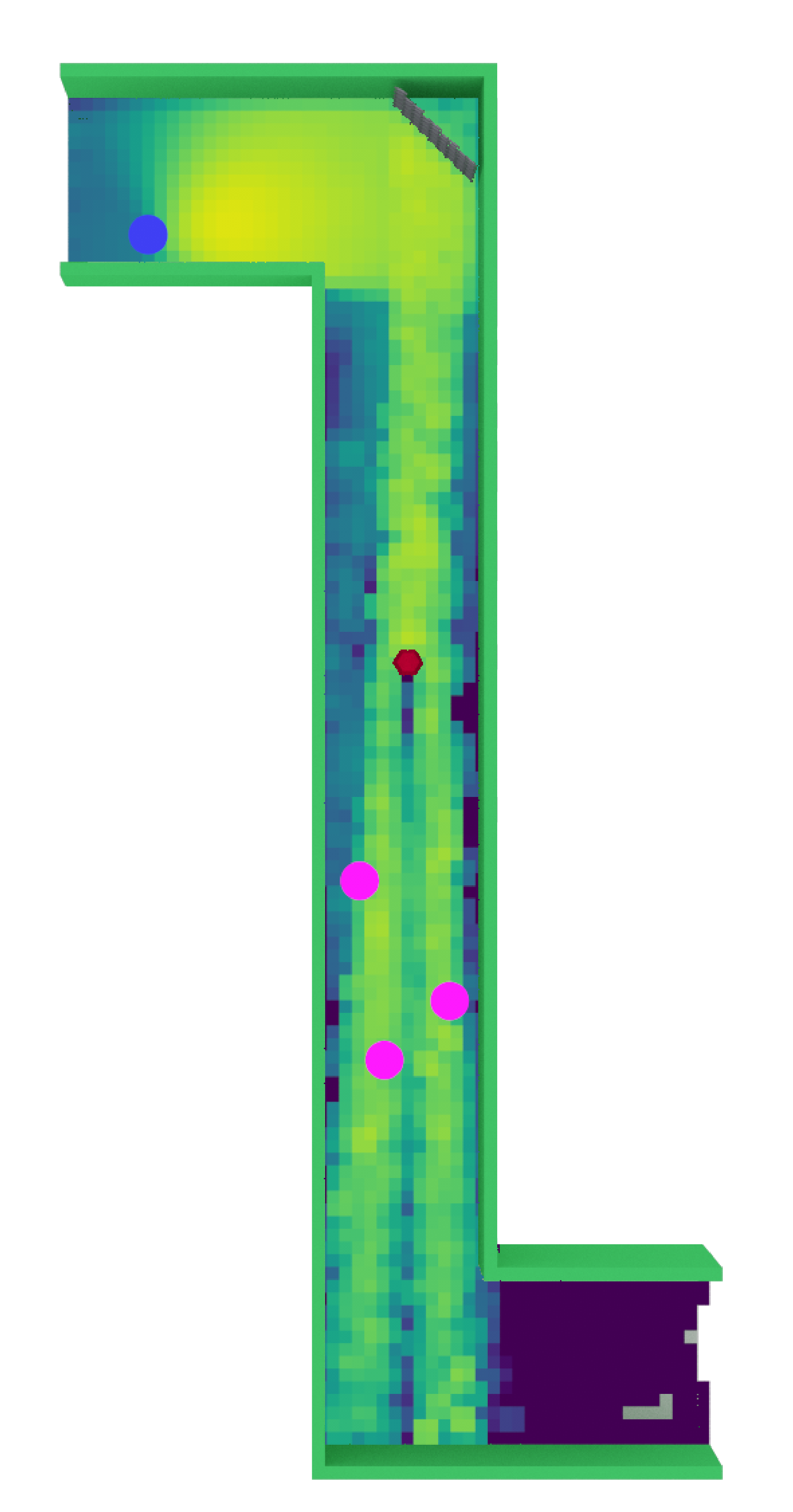}%
  \label{Figure:beamfocusing_ma}}
  \hfill
  \subfloat{\includegraphics[width=0.3\linewidth]{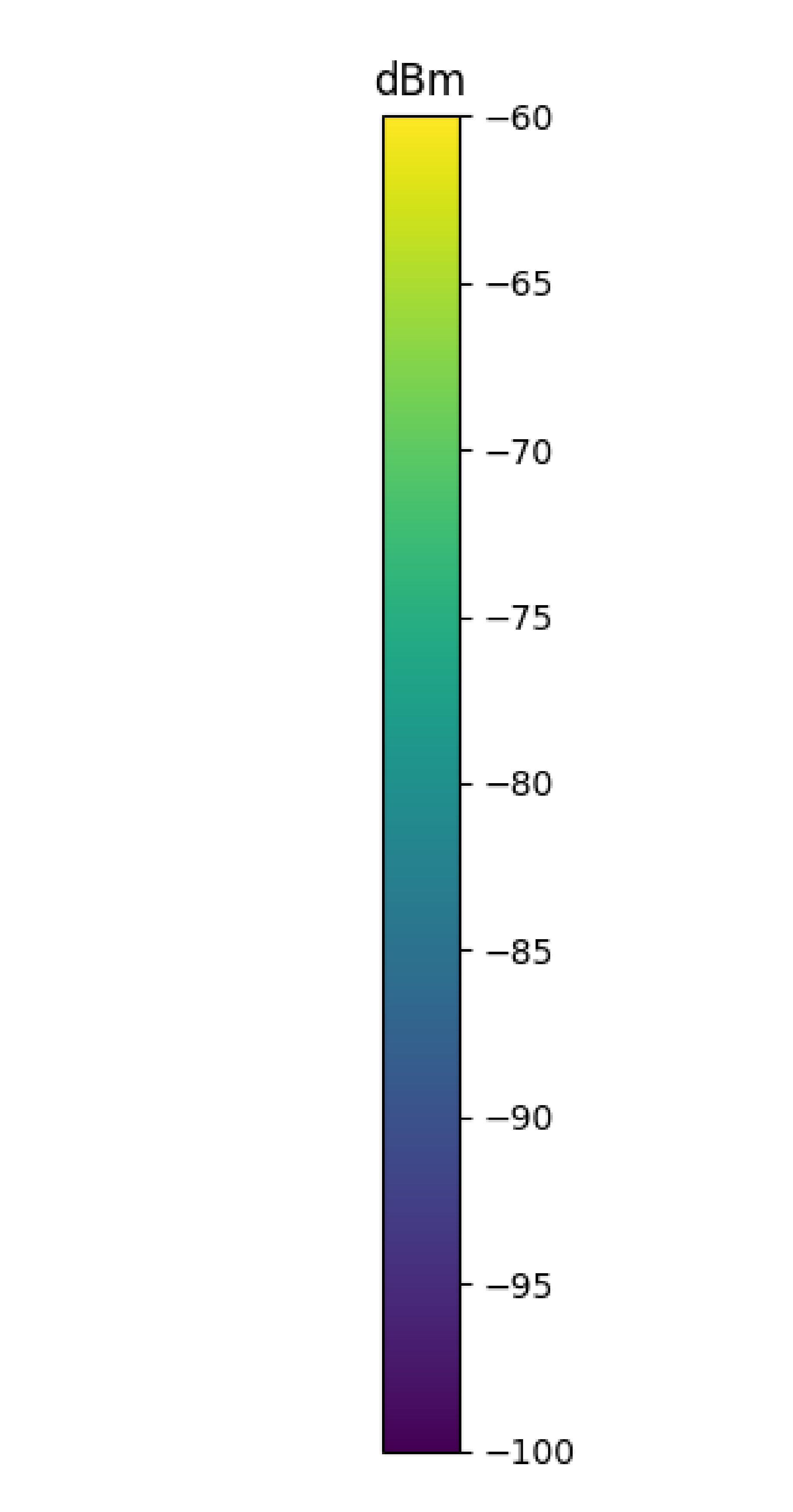}%
  \label{Figure:scale}}
  \caption{{Spatial RSSI distribution comparison across different control approaches in hallway environment during evaluation. Heat maps demonstrate progressive performance improvement from (a) no reflector (-108.2 dBm) and (b) flat reflector (-90.1 dBm) to reinforcement learning methods: (c) single-agent beam-focusing (-74.22 dBm), (d) multi-agent column-based (-72.71 dBm), and (e) multi-agent beam-focusing (-69.4 dBm). Color scale indicates RSSI values in dBm, with user positions marked as purple circles.}}
  \label{Figure:baseline_comparison}
\end{figure}

\begin{figure}[!ht]
    \centering
    \captionsetup{justification=centering}
    \includegraphics[width=1\linewidth]{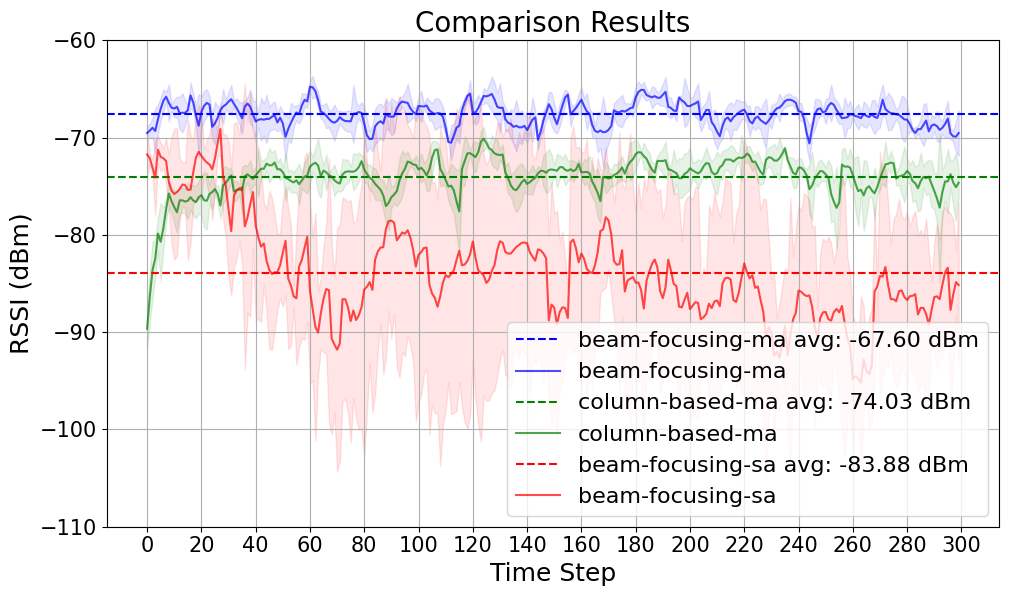}
    \caption{{Temporal RSSI performance comparison of reinforcement learning approaches over 300 simulation steps during evaluation. Multi-agent beam-focusing achieves superior performance (-67.60 dBm average), outperforming column-based multi-agent control (-74.03 dBm) and single-agent beam-focusing (-79.30 dBm). Solid lines represent mean values, shaded regions indicate ±1 standard deviation, and horizontal dashed lines show temporal averages.}}
    \label{Figure:baseline_eval}
\end{figure}

{
To comprehensively assess the effectiveness of the proposed MARL framework, we evaluate the trained policies against multiple baseline approaches and alternative control mechanisms. The comparative analysis encompasses scenarios ranging from passive configurations (no reflector, static reflector) to alternative algorithmic implementations (single-agent reinforcement learning, column-based control), providing comprehensive performance benchmarking across different system complexity levels.

The evaluation employs dynamic user mobility patterns where user locations are updated every four simulation steps to emulate realistic movement in hallway environments. The mobility pattern challenges the adaptive capabilities of different control mechanisms while providing consistent evaluation conditions across all methods. The comparison includes five distinct approaches: no reflector baseline, flat static reflector, single-agent beam-focusing (beam-focusing-sa), multi-agent column-based control (column-based-ma), and the proposed multi-agent beam-focusing approach (beam-focusing-ma).

For fair comparison, all reinforcement learning methods are provided with a brief adaptation period of three simulation steps following each user location change, allowing the algorithms to adjust to new environmental conditions. The adaptation window reflects realistic deployment scenarios where brief reconfiguration delays (one simulation step in our work) are acceptable while maintaining continuous service quality.

Fig. \ref{Figure:baseline_comparison} presents the spatial RSSI distribution and average performance across different approaches. The visualization reveals dramatic performance improvements when transitioning from passive to intelligent control mechanisms. The no reflector scenario achieves -108.2 dBm average RSSI, representing the baseline propagation condition without any signal enhancement. The flat reflector configuration provides modest improvement to -90.1 dBm, demonstrating the fundamental benefit of passive reflection while highlighting the limitations of non-adaptive approaches.

The spatial heat maps illustrate how intelligent control mechanisms create focused signal enhancement regions corresponding to user locations. The beam-focusing approaches demonstrate superior spatial selectivity, with concentrated high-signal regions at user positions, while the column-based method shows more distributed improvement patterns due to the constraint of shared azimuth angles within columns.

The reinforcement learning approaches demonstrate substantial performance advantages over static configurations, with all adaptive methods achieving RSSI improvements exceeding 15 dB compared to flat reflector baselines. The proposed beam-focusing-ma approach achieves the highest performance at -69.4 dBm average RSSI, representing a 38.8 dB improvement over the no reflector baseline and a 20.7 dB improvement over the static flat reflector.

The performance hierarchy clearly emerges from the evaluation: beam-focusing-ma (-69.4 dBm) $>$ column-based-ma (-72.71 dBm) $>$ beam-focusing-sa (-74.22 dBm). The ranking validates both the effectiveness of multi-agent coordination and the importance of unconstrained beam-focusing capabilities for optimal performance.

The single-agent approach with beam-focusing (beam-focusing-sa) achieves -74.22 dBm, demonstrating the challenges of learning complex coordination patterns from a single reward signal. While still providing significant improvement over static methods, the 4.82 dB performance gap compared to the multi-agent approach illustrates the value of task decomposition and specialized agent training.

Fig. \ref{Figure:baseline_eval} presents the temporal evolution of RSSI performance across 300 simulation steps, revealing dynamic adaptation characteristics of different control mechanisms. The beam-focusing-ma approach maintains consistently superior performance with an average RSSI of -67.60 dBm across all time steps, demonstrating stable adaptation to user mobility patterns.

The temporal analysis reveals important adaptation dynamics following user movement events. The beam-focusing-ma method exhibits rapid recovery from mobility-induced performance dips, typically requiring only one simulation step to restore optimal performance levels. The rapid adaptation capability stems from the effective coordination learned during training and the responsive nature of the CTDE framework.

The column-based-ma approach achieves -74.03 dBm average performance, representing a 6.43 dB reduction compared to unrestricted beam-focusing. The performance gap quantifies the cost of hardware simplification, where the reduction in servo requirements comes at the expense of spatial control flexibility. Despite this limitation, the column-based approach still provides substantial improvement over single-agent methods. Therefore, applications requiring maximum performance benefit from the full beam-focusing approach, while cost-sensitive deployments may accept the 6.43 dB performance reduction in exchange for simplified hardware implementation.

The shaded regions in Fig. \ref{Figure:baseline_eval} reveal performance variance characteristics across different approaches. The beam-focusing-ma method demonstrates the lowest variance, indicating consistent performance across different user configurations and mobility patterns. The stability reflects the robustness of the multi-agent coordination strategies and the effectiveness of the training methodology in producing reliable policies.

The single-agent approach exhibits higher performance variance, particularly during adaptation periods following user movement. The -79.30 dBm average RSSI for beam-focusing-sa, while still providing significant improvement over static methods, demonstrates the challenges of comprehensive system optimization through single-agent learning.
}

\subsection{Multi-User and Scalability Evaluation}

{
To assess the generalizability and scalability of the proposed MARL framework, we evaluate the trained multi-agent policies across scenarios with varying numbers of users and reflector configurations. Each user is allocated multiple columns with uniform spacing. For instance, with three users, user 1 is assigned columns 1, 4, and 7; user 2 is assigned columns 2, 5, and 8; and so forth. The evaluation employs policies pretrained exclusively on three-user scenarios to control reflectors in two-, three-, and four-user configurations, as well as reflector arrays ranging from 5 rows (45 tiles) to 11 rows (99 tiles). User locations are dynamically updated every four simulation steps to emulate realistic mobility patterns in hallway environments.
}

\subsubsection{Multi-User Adaptability Analysis}

\begin{figure}[!t]
    \centering
    \captionsetup{justification=centering}
    \includegraphics[width=1\linewidth]{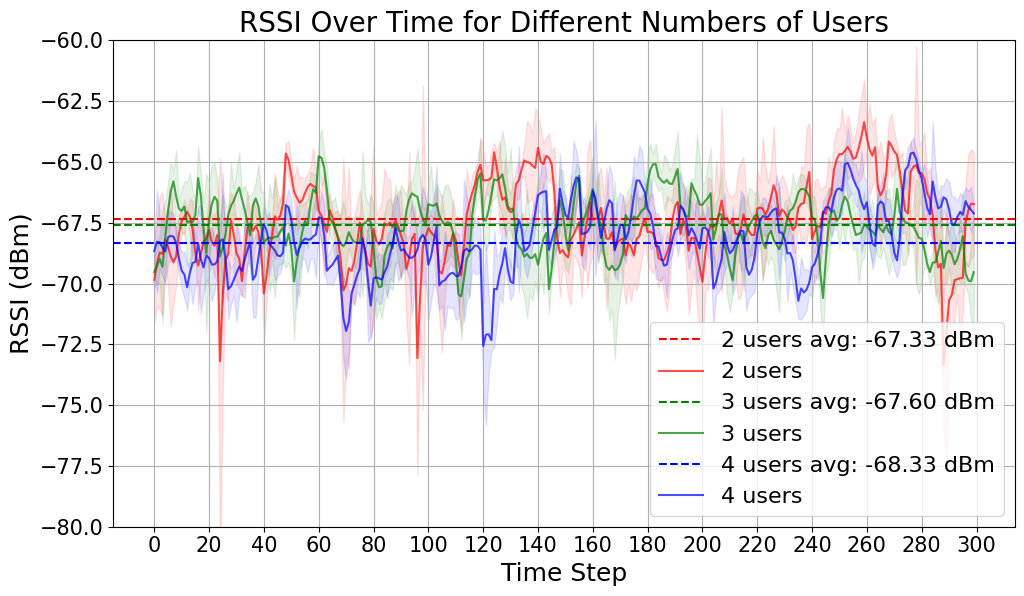}
    \caption{{Performance evaluation of multi-agent beam-focusing control across varying user densities. RSSI measurements over time demonstrate system adaptability with average performance of -67.33 dBm (2 users), -67.60 dBm (3 users), and -68.33 dBm (4 users). Solid lines show mean values, shaded regions represent ±1 standard deviation, and horizontal dashed lines indicate temporal averages.}}
    \label{Figure:multiuser_eval}
\end{figure}

{
Fig. \ref{Figure:multiuser_eval} presents the RSSI performance over time for different user numbers. The results demonstrate that the MARL-controlled reflector consistently improves RSSI performance across all user configurations, with average RSSI values of -67.33 dBm (2 users), -67.60 dBm (3 users), and -68.33 dBm (4 users). The improvement observed with fewer users can be attributed to resource allocation dynamics, where reduced user count allows more reflecting elements to be dedicated to each individual user, resulting in enhanced beam focusing and signal strength.

The temporal behavior exhibits characteristic performance patterns with periodic fluctuations in RSSI values aligning with user movement events occurring at four-step intervals. The system achieves rapid adaptation, generally requiring only a single simulation step to recalibrate the reflector setup and reestablish optimal beam steering toward updated user positions. During periods when users experience blocked line-of-sight conditions with the reflector, the system strategically configures selected reflecting elements to redirect signals via wall reflections, creating indirect propagation paths while maintaining RSSI performance above -72 dBm even under challenging propagation conditions.
}

\subsubsection{Reflector Size Scalability Analysis}

\begin{figure}[!t]
    \centering
    \captionsetup{justification=centering}
    \includegraphics[width=1\linewidth]{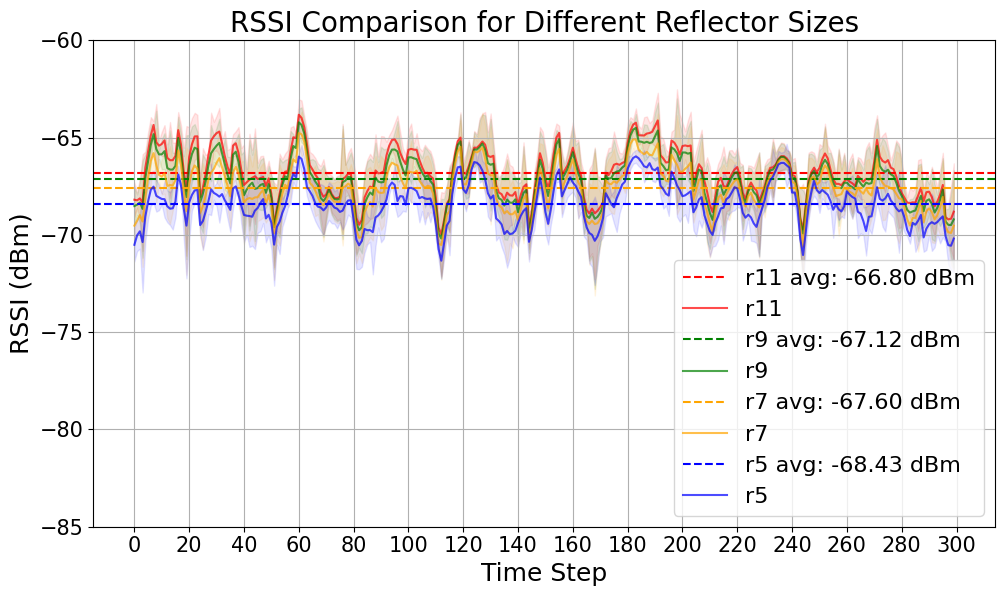}
    \caption{{Scalability evaluation of multi-agent beam-focusing control across varying reflector sizes. Average RSSI measurements demonstrate performance improvement with increased array size: -68.43 dBm (r5, 45 tiles), -67.60 dBm (r7, 63 tiles), -67.12 dBm (r9, 81 tiles), and -66.80 dBm (r11, 99 tiles). Solid lines show mean values, shaded regions represent ±1 standard deviation, and horizontal dashed lines indicate temporal averages.
}}
    \label{Figure:reflector_sizes_eval}
\end{figure}

{
Fig. \ref{Figure:reflector_sizes_eval} illustrates the temporal RSSI performance across different reflector sizes over 300 simulation steps, revealing consistent improvement in signal strength as reflector size increases. The 5-row configuration (r5) achieves -68.43 dBm average RSSI, improving to -67.60 dBm for the 7-row training configuration (r7), -67.08 dBm for 9 rows (r9), and -66.80 dBm for the largest 11-row configuration (r11). The performance improvement with increased reflector size can be attributed to enhanced spatial degrees of freedom and beam-focusing capabilities, with larger arrays providing more controllable elements for finer spatial resolution in beam steering.

The successful generalization of policies trained on the r7 configuration to both smaller and larger reflector arrays highlights a key advantage of the MARL approach. The learned coordination strategies scale effectively across different hardware configurations without requiring retraining, indicating that fundamental multi-element coordination principles captured during training remain applicable across varying system scales. The diminishing returns observed between larger configurations suggest performance saturation effects, the improvement from r5 to r7 (0.76 dBm) exceeds that from r7 to r9 (0.47 dBm), with the smallest gain from r9 to r11 (0.30 dBm), indicating that the reflector cannot enhance signals beyond the power limitations of the access point.
}

\subsubsection{Scalability Implications}

{
The performance variability across both user density and reflector size evaluations demonstrates remarkable consistency, suggesting robust policy operation independent of specific array dimensions or user configurations. The stability indicates that learned coordination strategies effectively scale across varying hardware specifications while maintaining predictable performance characteristics. These results validate that the MARL framework provides effective scalability for reflector systems, enabling deployment flexibility based on performance requirements and hardware constraints. The consistent performance improvement with increased reflector size confirms theoretical expectations that larger arrays provide enhanced beam-focusing capabilities, while the policy generalization capability reduces deployment complexity across different hardware specifications.
}

\subsection{System Robustness Analysis}

{
To evaluate the robustness of the proposed MARL framework to different system configurations and design choices, we conduct comprehensive analyses examining the impact of reflector element grouping strategies and reward function variations on system performance. These investigations assess the system's sensitivity to implementation decisions and validate the framework's reliability across diverse deployment scenarios.
}

\subsubsection{Reflector Element Grouping Pattern Analysis}

\begin{figure}[!t]
    \centering
    \captionsetup{justification=centering}
    \includegraphics[width=1\linewidth]{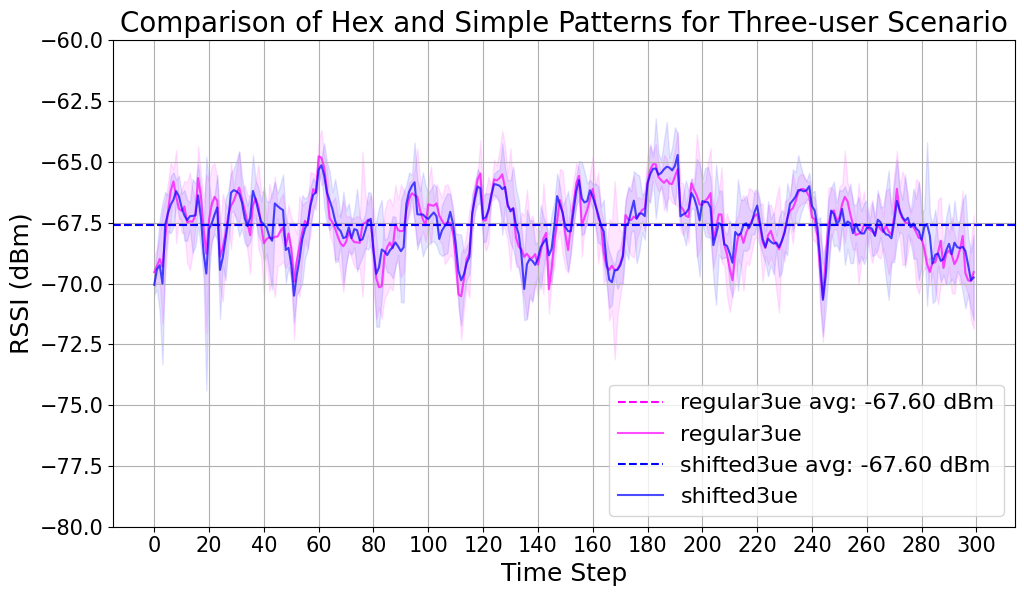}
    \caption{{Performance comparison of reflector element grouping strategies for three-user MARL control. RSSI measurements demonstrate equivalent performance between column-based grouping (regular3ue) and spatially distributed hexagonal pattern (shifted3ue), both achieving -67.60 dBm average RSSI. Solid lines show mean values, shaded regions represent ±1 standard deviation, and horizontal dashed lines indicate temporal averages.}}
    \label{Figure:pattern3ue_eval}
\end{figure}
{
We investigate two distinct grouping strategies for assigning reflecting elements to serve different users: conventional column-based grouping and a spatially distributed hexagonal pattern designed to maximize the effective aperture of each group. The conventional approach employs simple column-based grouping where consecutive columns are assigned to individual agents serving specific users. In contrast, the hexagonal or shifted pattern maximizes spatial diversity by distributing elements across the reflector surface. For a 7 × 9 reflector configuration, the shifted assignment algorithm initially assigns first row elements sequentially to groups 1 through 9, with subsequent rows undergoing a left shift operation by 4 positions. This arrangement ensures elements surrounding any central element belong to different groups, maintaining a minimum Manhattan distance of 4 between elements of the same group.

Fig. \ref{Figure:pattern3ue_eval} presents the temporal RSSI performance comparison between regular column-based grouping (regular3ue) and spatially distributed hexagonal pattern (shifted3ue) for a three-user scenario. Both configurations demonstrate similar performance characteristics, with average RSSI values of -67.60 dBm for both grouping strategies. The temporal evolution shows closely matched behavior across 300 simulation steps, with both patterns exhibiting similar variance and adaptation dynamics during user mobility events.
}

\subsubsection{Reward Function Sensitivity Analysis}

\begin{figure}[!t]
    \centering
    \captionsetup{justification=centering}
    \includegraphics[width=1\linewidth]{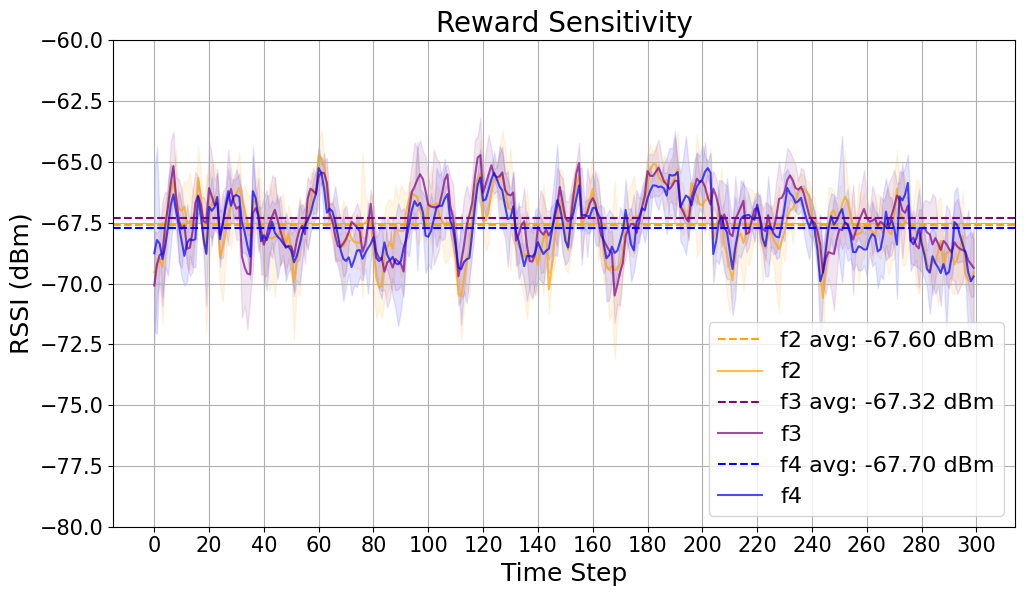}
    \caption{{Reward function sensitivity analysis for MARL-controlled intelligent reflecting surface across different distance normalization factors. RSSI measurements demonstrate robustness to reward engineering with minimal performance variation: -67.60 dBm (f2, $\mathbf{d^2}$ factor), -67.32 dBm (f3, $\mathbf{d^3}$ factor), and -67.70 dBm (f4, $\mathbf{d^4}$ factor). Solid lines show mean values, shaded regions represent ±1 standard deviation, and horizontal dashed lines indicate temporal averages.}}
    \label{Figure:rew_eval}
\end{figure}

{
We examine the impact of distance-based reward normalization on system performance by modifying the primary reward function that utilizes average RSSI across all users. The modifications attempt to remove received power dependency on distance according to the path loss model $P_{rx} = P_{tx} * \big{(}\nicefrac{\lambda_s}{4 \pi d}\big{)}^n$, where $P_{rx}$ represents received power, $P_{tx}$ denotes transmit power, $\lambda_s$ is the signal wavelength, $d$ is the distance, and $n$ is the path loss exponent dependent on the environment. We multiply measured RSSI of each user by factor $d^n$, creating distance-normalized reward variants with different path loss exponents (f2 through f4 representing $n = 2$ through $n = 4$).

Fig. \ref{Figure:rew_eval} presents temporal RSSI performance across different distance normalization factors. The distance factor multiplier exhibits minimal impact on reflector performance, with average RSSI values clustering tightly around -67.60 dBm (f2), -67.32 dBm (f3), and -67.70 dBm (f4). Performance variations remain within approximately 0.3 dBm across all normalization schemes, indicating robust learning behavior despite reward function modifications.
}

\subsubsection{Robustness Implications}

{
The performance equivalence between grouping strategies indicates that the MARL framework demonstrates robustness to reflector element organization patterns. The DRL algorithm successfully adapts to different spatial arrangements without requiring modifications to training procedures or policy architecture.

The invariance to reward scaling can be explained through mathematical properties of the PPO algorithm and advantage function normalization. The clipped surrogate objective, $L^{CLIP}$, depends fundamentally on advantage function estimates, which undergo normalization during training by subtracting the mean and dividing by the standard deviation within each minibatch, effectively making the learning process less sensitive to absolute reward value scales \cite{engstrom2020implementation}. The normalization ensures policy gradient direction remains consistent regardless of reward magnitude, aligning with theoretical work on reward invariance in policy gradient methods \cite{ibrahim2024comprehensive, ciosek2020expected}.

These findings provide significant flexibility for practical system implementation. System designers can prioritize mechanical simplicity, manufacturing constraints, or maintenance accessibility when determining reflector element grouping without compromising performance. Similarly, the robustness to reward engineering choices provides practical advantages for deployment scenarios where precise reward function design may be challenging, as the MARL system demonstrates stable learning behavior across different reward formulations.
}

\subsection{Position Information Noise Robustness Analysis}

\begin{figure}[!t]
    \centering
    \captionsetup{justification=centering}
    \includegraphics[width=1\linewidth]{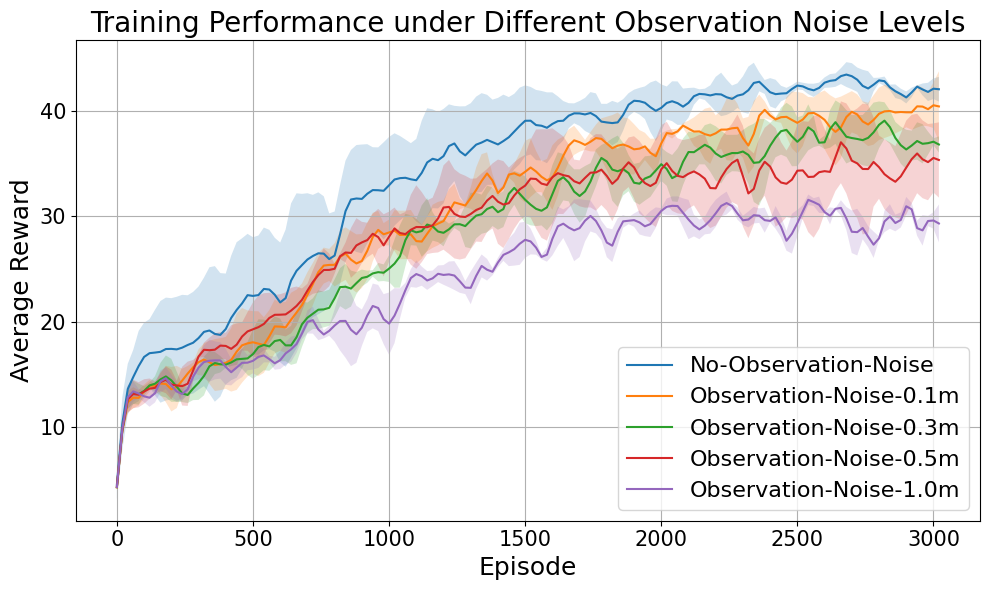}
    \caption{{Training performance under different user positioning noise levels. Solid lines represent mean training rewards with shaded regions indicating variance across multiple training runs.}}
    \label{Figure:training_noise}
\end{figure}

\begin{figure}[!t]
    \centering
    \captionsetup{justification=centering}
    \includegraphics[width=1\linewidth]{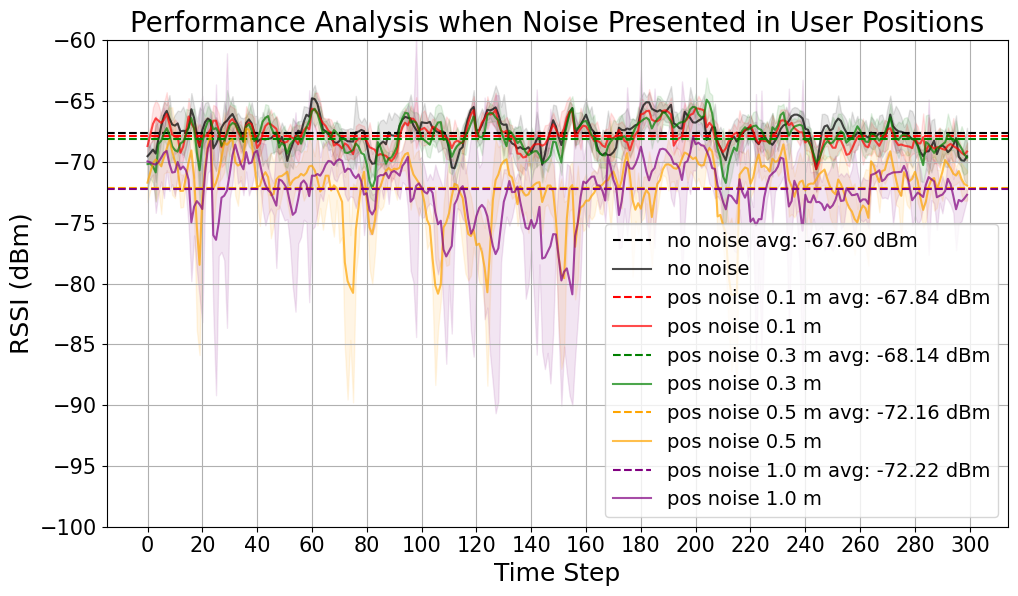}
    \caption{{Performance analysis when noise is present in user positions during evaluation. Solid lines show mean values, shaded regions represent ±1 standard deviation, and horizontal dashed lines indicate temporal averages.}}
    \label{Figure:pos_noise_eval}
\end{figure}

{
Practical deployment scenarios inevitably involve imperfect user localization due to measurement uncertainties, sensor limitations, and environmental factors. To evaluate system resilience under realistic conditions, we conduct analysis of MARL performance degradation under varying degrees of positional noise. This investigation addresses a critical deployment consideration, as indoor positioning systems typically achieve accuracies within submeters depending on the localization technology employed.

Gaussian noise with zero mean and standard deviations of 0.1, 0.3, 0.5, and 1.0 meters is added independently to each spatial coordinate of user positions during both training and evaluation phases. The noise model represents realistic positioning uncertainties encountered in practical indoor environments using technologies such as WiFi fingerprinting, ultra-wideband (UWB), or vision-based localization systems. Training is conducted for 3000 episodes under each noise condition using identical hyperparameters as the baseline configuration, while evaluation employs dynamic user mobility with position updates every four simulation steps.

Fig. \ref{Figure:training_noise} illustrates training performance across different noise levels in user positions, revealing the algorithm's ability to maintain stable convergence despite imperfect positional information. The no-noise baseline achieves maximum average reward of approximately 42, while increasing noise levels demonstrate graceful degradation: 0.1m noise achieves 39, 0.3m achieves 36, 0.5m achieves 35, and 1.0m noise maintains 30 reward. Notably, all noise conditions exhibit consistent convergence behavior with similar learning curves, indicating robust policy development under uncertainty. The convergence characteristics remain stable across noise levels, with initial rapid learning phases completing within 1000 episodes and steady improvements continuing through episode 2000.

Fig. \ref{Figure:pos_noise_eval} presents temporal RSSI performance under positional noise during evaluation, demonstrating system adaptability to imperfect location information. Average performance degrades gradually with increasing noise: -67.44 dBm (0.1m), -68.14 dBm (0.3m), -72.16 dBm (0.5m), and -72.22 dBm (1.0m), compared to -67.60 dBm without noise. However, the system demonstrates consistent adaptation capabilities, with performance fluctuations remaining bounded and average RSSI values maintaining stable levels throughout the 300-step evaluation period.

The observed robustness stems from two fundamental design characteristics. First, the focal point control abstraction operates at a higher spatial level than individual tile control, inherently reducing sensitivity to precise positional accuracy requirements. The geometric relationship between focal points and user locations remains approximately valid despite moderate positioning errors. Second, the multi-agent coordination provides spatial diversity across reflecting elements, enabling the system to compensate for localization uncertainties through distributed optimization.

}

\section{CONCLUSION AND FUTURE WORK}
\label{sec:conclusion}

{
This work demonstrates the effectiveness of MARL for controlling adjustable reflector arrays in wireless communication systems. Through extensive ray-tracing simulations using NVIDIA Sionna, our MARL framework with CTDE achieved substantial performance improvements, with 20.7 dB improvement over static flat reflectors. The system successfully adapts to user mobility patterns, typically requiring only one simulation step to restore optimal performance following location changes, while demonstrating robust scalability across different array sizes and user densities. The approach eliminates the complex channel estimation requirements that challenge conventional RIS systems by operating through spatial intelligence rather than electromagnetic precision.

While our mechanically adjustable metallic reflector approach demonstrates substantial performance improvements and eliminates CSI estimation requirements, it introduces mechanical complexity trade-offs that warrant acknowledgment. The servo-actuated reflector elements require high positioning accuracy, regular maintenance, and consideration of environmental factors such as temperature variations, wind loading, and mechanical wear over extended operation periods.
Despite these limitations, the substantial performance gains and reduced computational complexity provide compelling advantages that position this approach as a practical alternative for intelligent wireless environments, where moderate mechanical complexity is acceptable in exchange for simplified signal processing and enhanced coverage performance.

Future research will prioritize experimental validation through prototype development to verify mechanical feasibility assumptions, integration with commercial positioning systems to validate noise robustness findings in real deployments, computational latency optimization, and investigation of larger-scale deployments in diverse environmental conditions. While substantial engineering challenges remain before practical deployment, the fundamental approach shows promise for creating adaptive communication environments that can respond autonomously to changing conditions while maintaining economic feasibility, contributing to the growing body of research on intelligent propagation environment control.
}

\section*{ACKNOWLEDGMENT}
This material is based upon work supported by the U.S. Department of Energy, Office of Science, Office of Advanced Scientific Computing Research, Early Career Research Program under Award Number DE-SC-0023957.


\bibliographystyle{IEEEtran}
\bibliography{IEEEabrv,./main_ref}

\begin{thebibliography}{10}
\providecommand{\url}[1]{#1}
\csname url@samestyle\endcsname
\providecommand{\newblock}{\relax}
\providecommand{\bibinfo}[2]{#2}
\providecommand{\BIBentrySTDinterwordspacing}{\spaceskip=0pt\relax}
\providecommand{\BIBentryALTinterwordstretchfactor}{4}
\providecommand{\BIBentryALTinterwordspacing}{\spaceskip=\fontdimen2\font plus
\BIBentryALTinterwordstretchfactor\fontdimen3\font minus \fontdimen4\font\relax}
\providecommand{\BIBforeignlanguage}[2]{{%
\expandafter\ifx\csname l@#1\endcsname\relax
\typeout{** WARNING: IEEEtran.bst: No hyphenation pattern has been}%
\typeout{** loaded for the language `#1'. Using the pattern for}%
\typeout{** the default language instead.}%
\else
\language=\csname l@#1\endcsname
\fi
#2}}
\providecommand{\BIBdecl}{\relax}
\BIBdecl

\bibitem{direnzo:2020}
M.~Di~Renzo, A.~Zappone, M.~Debbah, M.-S. Alouini, C.~Yuen, J.~de~Rosny, and S.~Tretyakov, ``{Smart Radio Environments Empowered by Reconfigurable Intelligent Surfaces: How It Works, State of Research, and The Road Ahead},'' \emph{{IEEE Journal on Selected Areas in Communications}}, vol.~38, no.~11, pp. 2450--2525, 2020.

\bibitem{bjornson2022reconfigurable}
E.~Björnson, H.~Wymeersch, B.~Matthiesen, P.~Popovski, L.~Sanguinetti, and E.~de~Carvalho, ``{Reconfigurable Intelligent Surfaces: A Signal Processing Perspective with Wireless Applications},'' \emph{{IEEE Signal Processing Magazine}}, vol.~39, no.~2, pp. 135--158, 2022.

\bibitem{pan2022overview}
C.~Pan, G.~Zhou, K.~Zhi, S.~Hong, T.~Wu, Y.~Pan, H.~Ren, M.~D. Renzo, A.~Lee~Swindlehurst, R.~Zhang, and A.~Y. Zhang, ``{An Overview of Signal Processing Techniques for RIS/IRS-Aided Wireless Systems},'' \emph{{IEEE Journal of Selected Topics in Signal Processing}}, vol.~16, no.~5, pp. 883--917, 2022.

\bibitem{kim2022practical}
S.~Kim, H.~Lee, J.~Cha, S.-J. Kim, J.~Park, and J.~Choi, ``{Practical Channel Estimation and Phase Shift Design for Intelligent Reflecting Surface Empowered MIMO Systems},'' \emph{{IEEE Transactions on Wireless Communications}}, vol.~21, no.~8, pp. 6226--6241, 2022.

\bibitem{a9864655}
G.~C. Alexandropoulos, K.~Stylianopoulos, C.~Huang, C.~Yuen, M.~Bennis, and M.~Debbah, ``{Pervasive Machine Learning for Smart Radio Environments Enabled by Reconfigurable Intelligent Surfaces},'' \emph{{Proceedings of the IEEE}}, vol. 110, no.~9, pp. 1494--1525, 2022.

\bibitem{nasari2022benchmarking}
\BIBentryALTinterwordspacing
A.~Nasari, H.~Le, R.~Lawrence, Z.~He, X.~Yang, M.~Krell, A.~Tsyplikhin, M.~Tatineni, T.~Cockerill, L.~Perez, D.~Chakravorty, and H.~Liu, ``{Benchmarking the Performance of Accelerators on National Cyberinfrastructure Resources for Artificial Intelligence / Machine Learning Workloads},'' in \emph{{Practice and Experience in Advanced Research Computing 2022: Revolutionary: Computing, Connections, You}}, ser. PEARC '22.\hskip 1em plus 0.5em minus 0.4em\relax New York, NY, USA: Association for Computing Machinery, 2022. [Online]. Available: \url{https://doi.org/10.1145/3491418.3530772}
\BIBentrySTDinterwordspacing

\bibitem{le2024insight}
\BIBentryALTinterwordspacing
H.~Le, Z.~He, M.~Le, D.~Chakravorty, L.~M. Perez, A.~Chilumuru, Y.~Yao, and J.~Chen, ``{Insight Gained from Migrating a Machine Learning Model to Intelligence Processing Units},'' in \emph{{Practice and Experience in Advanced Research Computing 2024: Human Powered Computing}}, ser. PEARC '24.\hskip 1em plus 0.5em minus 0.4em\relax New York, NY, USA: Association for Computing Machinery, 2024. [Online]. Available: \url{https://doi.org/10.1145/3626203.3670527}
\BIBentrySTDinterwordspacing

\bibitem{qualcomm2024unlocking}
\BIBentryALTinterwordspacing
{Qualcomm}. (2024) {Unlocking On-device Generative AI with an NPU and Heterogeneous Computing}. [Online]. Available: \url{https://www.qualcomm.com/content/dam/qcomm-martech/dm-assets/documents/Unlocking-on-device-generative-AI-with-an-NPU-and-heterogeneous-computing.pdf}
\BIBentrySTDinterwordspacing

\bibitem{zahra:2021}
\BIBentryALTinterwordspacing
S.~Zahra, L.~Ma, W.~Wang, J.~Li, D.~Chen, Y.~Liu, Y.~Zhou, N.~Li, Y.~Huang, and G.~Wen, ``{Electromagnetic Metasurfaces and Reconfigurable Metasurfaces: A Review},'' \emph{{Frontiers in Physics}}, vol.~8, 2021. [Online]. Available: \url{https://www.frontiersin.org/articles/10.3389/fphy.2020.593411}
\BIBentrySTDinterwordspacing

\bibitem{basharat:2022}
\BIBentryALTinterwordspacing
S.~Basharat, M.~Khan, M.~Iqbal, U.~S. Hashmi, S.~A.~R. Zaidi, and I.~Robertson, ``{Exploring Reconfigurable intelligent surfaces for 6G: State-of-the-art and the road ahead},'' \emph{{IET Communications}}, vol.~16, no.~13, pp. 1458--1474, 2022. [Online]. Available: \url{https://ietresearch.onlinelibrary.wiley.com/doi/abs/10.1049/cmu2.12364}
\BIBentrySTDinterwordspacing

\bibitem{a9400843}
C.~Hu, L.~Dai, S.~Han, and X.~Wang, ``{Two-Timescale Channel Estimation for Reconfigurable Intelligent Surface Aided Wireless Communications},'' \emph{{IEEE Transactions on Communications}}, vol.~69, no.~11, pp. 7736--7747, 2021.

\bibitem{a10053657}
J.~Chen, Y.-C. Liang, H.~V. Cheng, and W.~Yu, ``{Channel Estimation for Reconfigurable Intelligent Surface Aided Multi-User mmWave MIMO Systems},'' \emph{{IEEE Transactions on Wireless Communications}}, vol.~22, no.~10, pp. 6853--6869, 2023.

\bibitem{a9328501}
X.~Wei, D.~Shen, and L.~Dai, ``{Channel Estimation for RIS Assisted Wireless Communications—Part I: Fundamentals, Solutions, and Future Opportunities},'' \emph{{IEEE Communications Letters}}, vol.~25, no.~5, pp. 1398--1402, 2021.

\bibitem{a10016718}
B.~Shamasundar, N.~Daryanavardan, and A.~Nosratinia, ``{Channel Training \& Estimation for Reconfigurable Intelligent Surfaces: Exposition of Principles, Approaches, and Open Problems},'' \emph{{IEEE Access}}, vol.~11, pp. 6717--6734, 2023.

\bibitem{a10666709}
A.~Abrardo, ``{Optimizing Reconfigurable Intelligent Surfaces in Multi-User Environments: A Multiport Network Theory Approach Leveraging Statistical CSI},'' \emph{{IEEE Transactions on Communications}}, vol.~73, no.~3, pp. 2047--2060, 2025.

\bibitem{lai2023blind}
W.~Lai, W.~Wang, F.~Xu, X.~Li, S.~Niu, and K.~Shen, ``{Blind beamforming for intelligent reflecting surface in fading channels without CSI},'' \emph{arXiv e-prints}, pp. arXiv--2305, 2023.

\bibitem{a9952197}
J.~An, C.~Xu, Q.~Wu, D.~W.~K. Ng, M.~Di~Renzo, C.~Yuen, and L.~Hanzo, ``{Codebook-Based Solutions for Reconfigurable Intelligent Surfaces and Their Open Challenges},'' \emph{{IEEE Wireless Communications}}, vol.~31, no.~2, pp. 134--141, 2024.

\bibitem{nazar2024revolutionizing}
A.~M. Nazar, M.~Y. Selim, and D.~Qiao, ``{Revolutionizing RIS Networks: Lidar-based Data-driven Approach to Enhance RIS Beamforming},'' \emph{{Sensors}}, vol.~25, no.~1, p.~75, 2024.

\bibitem{huang2020reconfigurable}
C.~Huang, R.~Mo, and C.~Yuen, ``{Reconfigurable Intelligent Surface Assisted Multiuser MISO Systems Exploiting Deep Reinforcement Learning},'' \emph{{IEEE Journal on Selected Areas in Communications}}, vol.~38, no.~8, pp. 1839--1850, 2020.

\bibitem{taha2020deep}
A.~Taha, Y.~Zhang, F.~B. Mismar, and A.~Alkhateeb, ``{Deep Reinforcement Learning for Intelligent Reflecting Surfaces: Towards Standalone Operation},'' in \emph{{2020 IEEE 21st International Workshop on Signal Processing Advances in Wireless Communications (SPAWC)}}, 2020, pp. 1--5.

\bibitem{taha2021enabling}
A.~Taha, M.~Alrabeiah, and A.~Alkhateeb, ``{Enabling Large Intelligent Surfaces With Compressive Sensing and Deep Learning},'' \emph{{IEEE Access}}, vol.~9, pp. 44\,304--44\,321, 2021.

\bibitem{choi2024deep}
H.~Choi, L.~V. Nguyen, J.~Choi, and A.~L. Swindlehurst, ``{A Deep Reinforcement Learning Approach for Autonomous Reconfigurable Intelligent Surfaces},'' in \emph{{2024 IEEE International Conference on Communications Workshops (ICC Workshops)}}, 2024, pp. 208--213.

\bibitem{sheen2021deep}
B.~Sheen, J.~Yang, X.~Feng, and M.~M.~U. Chowdhury, ``{A Deep Learning Based Modeling of Reconfigurable Intelligent Surface Assisted Wireless Communications for Phase Shift Configuration},'' \emph{{IEEE Open Journal of the Communications Society}}, vol.~2, pp. 262--272, 2021.

\bibitem{a10060056}
A.~Abdallah, A.~Celik, M.~M. Mansour, and A.~M. Eltawil, ``{Deep Reinforcement Learning Based Beamforming Codebook Design for RIS-aided mmWave Systems},'' in \emph{{2023 IEEE 20th Consumer Communications and Networking Conference (CCNC)}}, 2023, pp. 1020--1026.

\bibitem{a10758034}
P.-H. Liao, L.-H. Shen, P.-C. Wu, and K.-T. Feng, ``{Multi-Agent Deep Reinforcement Learning for Energy Efficient Multi-Hop STAR-RIS-Assisted Transmissions},'' in \emph{{2024 IEEE 100th Vehicular Technology Conference (VTC2024-Fall)}}, 2024, pp. 1--5.

\bibitem{busoniu2008comprehensive}
L.~Busoniu, R.~Babuska, and B.~De~Schutter, ``{A Comprehensive Survey of Mmultiagent Reinforcement Learning},'' \emph{{IEEE Transactions on Systems, Man, and Cybernetics, Part C (Applications and Reviews)}}, vol.~38, no.~2, pp. 156--172, 2008.

\bibitem{zhang2018fully}
K.~Zhang, Z.~Yang, H.~Liu, T.~Zhang, and T.~Basar, ``{Fully Decentralized Multi-agent Reinforcement Learning with Networked Agents},'' in \emph{{International Conference on Machine Learning}}.\hskip 1em plus 0.5em minus 0.4em\relax {PMLR}, 2018, pp. 5872--5881.

\bibitem{kar2013cal}
S.~Kar, J.~M. Moura, and H.~V. Poor, ``{QD-Learning: A Collaborative Distributed Strategy for Multi-Agent Reinforcement Learning Through Consensus+Innovations},'' \emph{{IEEE Transactions on Signal Processing}}, vol.~61, no.~7, pp. 1848--1862, 2013.

\bibitem{littman1994markov}
M.~L. Littman, ``{Markov Games as a Framework for Multi-agent Reinforcement Learning},'' in \emph{{Machine Learning Proceedings 1994}}.\hskip 1em plus 0.5em minus 0.4em\relax Elsevier, 1994, pp. 157--163.

\bibitem{shapley1953stochastic}
L.~S. Shapley, ``{Stochastic Games},'' \emph{{Proceedings of the National Academy of Sciences}}, vol.~39, no.~10, pp. 1095--1100, 1953.

\bibitem{park2023multi}
C.~Park, K.~Zhang, and A.~Ozdaglar, ``{Multi-player Zero-sum Markov Games with Networked Separable Interactions},'' \emph{{Advances in Neural Information Processing Systems}}, vol.~36, pp. 37\,354--37\,369, 2023.

\bibitem{hu2003nash}
J.~Hu and M.~P. Wellman, ``{Nash Q-learning for General-sum Stochastic Games},'' \emph{{Journal of Machine Learning Research}}, vol.~4, no. Nov, pp. 1039--1069, 2003.

\bibitem{lowe2017multi}
R.~Lowe, Y.~I. Wu, A.~Tamar, J.~Harb, O.~Pieter~Abbeel, and I.~Mordatch, ``{Multi-agent Actor-critic for Mixed Cooperative-Competitive Environments},'' \emph{{Advances in Neural Information Processing Systems}}, vol.~30, 2017.

\bibitem{a10261304}
B.~Hazarika, K.~Singh, S.~Biswas, S.~Mumtaz, and C.-P. Li, ``{Multi-Agent DRL-Based Task Offloading in Multiple RIS-Aided IoV Networks},'' \emph{{IEEE Transactions on Vehicular Technology}}, vol.~73, no.~1, pp. 1175--1190, 2024.

\bibitem{a10654286}
K.~Qi, Q.~Wu, P.~Fan, N.~Cheng, Q.~Fan, and J.~Wang, ``{Reconfigurable Intelligent Surface Assisted VEC Based on Multi-Agent Reinforcement Learning},'' \emph{{IEEE Communications Letters}}, vol.~28, no.~10, pp. 2427--2431, 2024.

\bibitem{a10508095}
Y.~Zhu, E.~Shi, Z.~Liu, J.~Zhang, and B.~Ai, ``{Multi-Agent Reinforcement Learning-Based Joint Precoding and Phase Shift Optimization for RIS-Aided Cell-Free Massive MIMO Systems},'' \emph{{IEEE Transactions on Vehicular Technology}}, vol.~73, no.~9, pp. 14\,015--14\,020, 2024.

\bibitem{a9314027}
G.~Zhou, C.~Pan, H.~Ren, K.~Wang, M.~Elkashlan, and M.~D. Renzo, ``{Stochastic Learning-Based Robust Beamforming Design for RIS-Aided Millimeter-Wave Systems in the Presence of Random Blockages},'' \emph{{IEEE Transactions on Vehicular Technology}}, vol.~70, no.~1, pp. 1057--1061, 2021.

\bibitem{a8972365}
W.~Khawaja, O.~Ozdemir, Y.~Yapici, F.~Erden, and I.~Guvenc, ``{Coverage Enhancement for NLOS mmWave Links Using Passive Reflectors},'' \emph{{IEEE Open Journal of the Communications Society}}, vol.~1, pp. 263--281, 2020.

\bibitem{a9500547}
A.~Deng, Y.~Liu, and D.~M. Blough, ``{Maximizing Coverage for mmWave WLANs with Dedicated Reflectors},'' in \emph{{ICC 2021 - IEEE International Conference on Communications}}, 2021, pp. 1--6.

\bibitem{le2024guiding}
H.~Le, O.~Bedir, M.~Ibrahim, J.~Tao, and S.~Ekin, ``{Guiding Wireless Signals with Arrays of Metallic Linear Fresnel Reflectors: A Low-cost, Frequency-versatile, and Practical Approach},'' in \emph{{2024 IEEE 100th Vehicular Technology Conference (VTC2024-Fall)}}, 2024, pp. 1--7.

\bibitem{a10279522}
Z.~Yu, C.~Feng, Y.~Zeng, T.~Li, and S.~Jin, ``{Wireless Communication Using Metal Reflectors: Reflection Modelling and Experimental Verification},'' in \emph{{ICC 2023 - IEEE International Conference on Communications}}, 2023, pp. 4701--4706.

\bibitem{sionna}
J.~Hoydis, S.~Cammerer, F.~{Ait Aoudia}, A.~Vem, N.~Binder, G.~Marcus, and A.~Keller, ``{Sionna: An Open-Source Library for Next-Generation Physical Layer Research},'' \emph{{arXiv preprint}}, Mar. 2022.

\bibitem{blender}
\BIBentryALTinterwordspacing
{Blender}. (2025) {The Blender Foundation}. {Blender}. [Online]. Available: \url{https://www.blender.org/}
\BIBentrySTDinterwordspacing

\bibitem{a10686755}
Y.~Wang, Z.~Liu, L.~Guo, and J.~Guo, ``{An Acceleration Method for Ray-Tracing in Modeling Reconfigurable Intelligent Surface Propagation},'' in \emph{{2024 IEEE International Symposium on Antennas and Propagation and INC/USNC‐URSI Radio Science Meeting (AP-S/INC-USNC-URSI)}}, 2024, pp. 2183--2184.

\bibitem{a10416965}
Z.~Yuan, J.~Zhang, V.~Degli-Esposti, Y.~Zhang, and W.~Fan, ``{Efficient Ray-Tracing Simulation for Near-Field Spatial Non-Stationary mmWave Massive MIMO Channel and Its Experimental Validation},'' \emph{{IEEE Transactions on Wireless Communications}}, vol.~23, no.~8, pp. 8910--8923, 2024.

\bibitem{rec_p2040}
\BIBentryALTinterwordspacing
I.~T. U.~R. Sector. (2024, Dec.) {Effects of building materials and structures on radiowave propagation above about 100MHz}. {International Telecommunications Union Radiocommunication Sector}. [Online]. Available: \url{https://www.itu.int/rec/R-REC-P.2040/en}
\BIBentrySTDinterwordspacing

\bibitem{schulman2017proximal}
J.~Schulman, F.~Wolski, P.~Dhariwal, A.~Radford, and O.~Klimov, ``{Proximal Policy Optimization Algorithms},'' \emph{{arXiv preprint arXiv:1707.06347}}, 2017.

\bibitem{yu2022surprising}
C.~Yu, A.~Velu, E.~Vinitsky, J.~Gao, Y.~Wang, A.~Bayen, and Y.~Wu, ``{The Surprising Effectiveness of PPO in Cooperative Multi-agent Games},'' \emph{{Advances in Neural Information Processing Systems}}, vol.~35, pp. 24\,611--24\,624, 2022.

\bibitem{a10663867}
J.~Queeney, I.~C. Paschalidis, and C.~G. Cassandras, ``{Generalized Policy Improvement Algorithms With Theoretically Supported Sample Reuse},'' \emph{{IEEE Transactions on Automatic Control}}, vol.~70, no.~2, pp. 1236--1243, 2025.

\bibitem{wang2020truly}
Y.~Wang, H.~He, and X.~Tan, ``{Truly Proximal Policy Optimization},'' in \emph{{Uncertainty in Artificial Intelligence}}.\hskip 1em plus 0.5em minus 0.4em\relax PMLR, 2020, pp. 113--122.

\bibitem{engstrom2020implementation}
\BIBentryALTinterwordspacing
L.~Engstrom, A.~Ilyas, S.~Santurkar, D.~Tsipras, F.~Janoos, L.~Rudolph, and A.~Madry, ``{Implementation Matters in Deep RL: A Case Study on PPO and TRPO},'' in \emph{International Conference on Learning Representations}, 2020. [Online]. Available: \url{https://openreview.net/forum?id=r1etN1rtPB}
\BIBentrySTDinterwordspacing

\bibitem{ibrahim2024comprehensive}
S.~Ibrahim, M.~Mostafa, A.~Jnadi, H.~Salloum, and P.~Osinenko, ``Comprehensive overview of reward engineering and shaping in advancing reinforcement learning applications,'' \emph{IEEE Access}, 2024.

\bibitem{ciosek2020expected}
K.~Ciosek and S.~Whiteson, ``{Expected Policy Gradients for Reinforcement Learning},'' \emph{Journal of Machine Learning Research}, vol.~21, no.~52, pp. 1--51, 2020.

\end{thebibliography}

\newpage

\vfill\pagebreak

\end{document}